\documentclass[final]{IEEEtran}
\usepackage[latin9]{inputenc}
\usepackage{setspace,mathdots} 
\usepackage{verbatim,url}
\usepackage{float}
\usepackage{wrapfig,multirow}
\usepackage{sidecap,stackrel}
\usepackage[T1]{fontenc}

\usepackage[usenames]{color}
\usepackage{prettyref,soul,xspace}
\usepackage{fancyhdr,lastpage}
\usepackage{graphicx,amsmath,amssymb,enumerate}

\usepackage{amsmath,amsfonts,amssymb,amsthm,graphicx,graphics,subfigure,epsfig,enumerate,bm}
\usepackage{ccaption}

\usepackage{algorithm}
\usepackage{algorithmic}

\newcommand{\yc}[1]{{\color{black} #1}}
\newcommand{\bp}{{\boldsymbol p}}
\newcommand{\bx}{{\boldsymbol x}}

\newcommand{\y}{{\boldsymbol y}}
\newcommand{\be}{{\boldsymbol e}}

\newcommand{\bXi}{{\boldsymbol \Xi}}

\newcommand{\bzeta}{{\boldsymbol \zeta}}

\newcommand{\argmax}{\mathop{\rm argmax}}
\newcommand{\argmin}{\mathop{\rm argmin}}

\newcommand{\bK}{{\boldsymbol K}} 
\newcommand{\bA}{{\boldsymbol A}}
\newcommand{\bB}{{\boldsymbol B}}
\newcommand{\bL}{{\boldsymbol L}}

\newcommand{\bZ}{{\boldsymbol Z}}
\newcommand{\bc}{{\boldsymbol c}}

\newcommand{\bh}{{\boldsymbol h}}

\newcommand{\bQ}{{\boldsymbol Q}}
\newcommand{\bR}{{\boldsymbol R}}
\newcommand{\bS}{{\boldsymbol S}}

\newcommand{\bb}{{\boldsymbol b}}
\newcommand{\by}{{\boldsymbol y}}
\newcommand{\bu}{{\boldsymbol u}}
\newcommand{\bg}{{\boldsymbol g}}
\newcommand{\bq}{{\boldsymbol q}}
\newcommand{\bv}{{\boldsymbol v}}

\newcommand{\bI}{{\boldsymbol I}}
\newcommand{\bY}{{\boldsymbol Y}}

\newcommand{\bX}{{\boldsymbol X}}

\newcommand{\bW}{{\boldsymbol W}}

\newcommand{\balpha}{{\boldsymbol \alpha}}
\newcommand{\bbeta}{{\boldsymbol \beta}}
\newcommand{\bPhi}{{\boldsymbol \Phi}}
\newcommand{\bGamma}{{\boldsymbol \Gamma}}
\newcommand{\bnu}{{\boldsymbol \nu}}

\newcommand{\kron}{\otimes}

\newcommand{\expectation}{{\mathbb{E}}}

\newrefformat{eq}{(\ref{#1})}
\newrefformat{sec}{Section~\ref{#1}}
\newrefformat{algo}{Algorithm~\ref{#1}}
\newrefformat{fig}{Fig.~\ref{#1}}
\newrefformat{tab}{Table~\ref{#1}}
\newrefformat{rmk}{Remark~\ref{#1}}
\newrefformat{clm}{Claim~\ref{#1}}
\newrefformat{def}{Definition~\ref{#1}}
\newrefformat{cor}{Corollary~\ref{#1}}
\newrefformat{lmm}{Lemma~\ref{#1}}
\newrefformat{lemma}{Lemma~\ref{#1}}
\newrefformat{prop}{Proposition~\ref{#1}}

\usepackage{algorithm, multirow}
\usepackage{algorithmic}
\usepackage{graphicx}
\usepackage{amsmath}
\usepackage{amssymb}
\usepackage{wrapfig}

\DeclareMathOperator{\trace}{Tr}
\DeclareMathOperator{\toep}{toep}

\DeclareMathOperator{\diag}{diag}

\newcommand{\ba}{{\boldsymbol a}} 
\newcommand{\bw}{{\boldsymbol w}} 

\newcommand{\cA}{{\mathcal A}}

\theoremstyle{definition}

\title{Guaranteed Blind Sparse Spikes Deconvolution via Lifting and Convex Optimization}

\author{Yuejie Chi,~\IEEEmembership{Member,~IEEE} 
\thanks{Copyright (c) 2014 IEEE. Personal use of this material is permitted. However, permission to use this material for any
other purposes must be obtained from the IEEE by sending a request to pubs-permissions@ieee.org.}
\thanks{The author is with Department of Electrical and Computer Engineering and Department of Biomedical Informatics, The Ohio State University, Columbus, Ohio 43210. This work is supported in part by ONR under grant N00014-15-1-2387 and NSF under grant CCF-1527456. Email: chi.97@osu.edu. }\thanks{Preliminary results of this paper were presented in part at the Asilomar Conference on Signals, Systems, and Computers
(Asilomar), Pacific Grove, CA, Nov. 2015.}}

\begin{document}
\theoremstyle{plain}\newtheorem{lemma}{\textbf{Lemma}}\newtheorem{corollary}{\textbf{Corollary}}\newtheorem{assumption}{\textbf{Assumption}}\newtheorem{prop}{\textbf{Proposition}}\newtheorem{theorem}{\textbf{Theorem}}
\newtheorem{example}{\textbf{Example}}
\theoremstyle{remark}\newtheorem{remark}{\textbf{Remark}}

\maketitle

\begin{abstract}
Neural recordings, returns from radars and sonars, images in astronomy and single-molecule microscopy can be modeled as a linear superposition of a small number of scaled and delayed copies of a band-limited or diffraction-limited point spread function, which is either determined by the nature or designed by the users; in other words, we observe the convolution between a point spread function and a sparse spike signal with unknown amplitudes and delays. While it is of great interest to accurately resolve the spike signal from as few samples as possible, however, when the point spread function is not known a priori, this problem is terribly ill-posed. This paper proposes a convex optimization framework to simultaneously estimate the point spread function as well as the spike signal, by mildly constraining the point spread function to lie in a known low-dimensional subspace. By applying the lifting trick, we obtain an underdetermined linear system of an ensemble of signals with joint spectral sparsity, to which atomic norm minimization is applied. Under mild randomness assumptions of the low-dimensional subspace as well as a separation condition of the spike signal, we prove the proposed algorithm, dubbed as AtomicLift, is guaranteed to recover the spike signal up to a scaling factor as soon as the number of samples is large enough. The extension of AtomicLift to handle noisy measurements is also discussed. Numerical examples are provided to validate the effectiveness of the proposed approaches.
\end{abstract}



\section{Introduction}

In many applications, the goal is to estimate the set of delays and amplitudes of point sources contained in a sparse spike signal $x(t)$ from its convolution with a band-limited or diffraction-limited point spread function (PSF) $g(t)$, which is either determined by the nature or designed by the users. This describes the problem of estimating target locations in radar and sonar, firing times of neurons, direction-of-arrivals in array signal processing, etc. 

When the PSF is assumed perfectly known, many algorithms have been developed to retrieve the spike signal, ranging from subspace methods such as MUSIC \cite{Schmidt1986MUSIC} and ESPRIT \cite{RoyKailathESPIRIT1989} to total variation minimization  \cite{candes2014towards}. However, in many applications, the PSF is not known a priori, and must be estimated together with the spike model, referred to as {\em blind spikes deconvolution}. As an example, in neural spike train decoding, the characteristic function of the neurons is determined by the nature and needs to be calibrated \cite{gerstein1964simultaneous}. As another example, in blind channel estimation for wireless communications, the transmitted signal is modulated by unknown data symbols, therefore the receiver has to perform joint channel estimation and symbol decoding. A related problem is {\em blind calibration} of uniform linear arrays \cite{Friedlander1988}, where it is desirable to calibrate the gains of the array antennas in a blind fashion. 

Broadly speaking, blind deconvolution of two signals from their convolution falls into the category of bilinear inverse problems,  which is in general ill-posed without further constraints. The identifiability, up to an unavoidable scaling ambiguity, of these problems has recently been investigated in \cite{li2015identifiability}\nocite{li2015unified}--\cite{choudhary2013identifiability} under the constraints that one or both signals are sparse or lie in some known subspace. Along the algorithmic line of research, conventional approaches for blind deconvolution are typically based on expectation maximization \cite{Friedlander1988,Viberg1994}, which often suffer from local minima and lack performance guarantees. Recently, Ahmed, Recht and Romberg developed a provably-correct algorithm for blind deconvolution by assuming both signals lie in some known low-dimensional subspaces \cite{ahmed2014blind} under certain conditions. The key in their approach is the so-called lifting trick that translates the problem into an under-determined linear system with respect to a lifted rank-one matrix, which can be exactly recovered using a nuclear norm minimization algorithm. Ling and Strohmer \cite{ling2015self} further extended this framework by allowing one of the signal to be sparse in a known dictionary, and applied $\ell_1$-norm minimization to the lifted sparse matrix, for which sufficient conditions for exact recovery are also provided. Finally, Lee et.~al. proposed an alternating minimization framework \cite{lee2015stability} to the case when both signals are sparse in a known dictionary, and established convergence guarantees from a near-optimal number of samples under some conditions.

\subsection{Our Contributions}
In this paper, we study the problem of blind spikes deconvolution, where we want to jointly estimate the PSF and the spike signal composed of a small number of delayed and scaled Dirac functions. Since it is more convenient to work in the Fourier domain, we start by sampling the Fourier transform of the convolution, giving rise to a measurement vector $\boldsymbol{y} = \bg \odot \bx  + \bw \in\mathbb{C}^{N}$, where $\odot$ denotes point-wise product, $\bg\in\mathbb{C}^N$ is the sampled Fourier transform of the PSF, $\bx\in\mathbb{C}^N$ is the sampled Fourier transform of the spike signal, which is a sum of $K$ complex sinusoids with frequencies determined by the corresponding delays, $K$ is the number of spikes, and $\bw\in\mathbb{C}^N$ is an additive noise term. Our problem is to recover the set of spikes contained in $\bx$ from the possibly noisy observation $\by$. 

Motivated by \cite{ahmed2014blind}, we assume that the PSF $\bg$ lies in a known low-dimensional subspace $\bB\in\mathbb{C}^{N\times L}$, i.e. $\bg=\bB\bh$, $\bh\in\mathbb{C}^L$, where the orientation of $\bg$ in the subspace, given by $\bh$, still needs to be estimated. This assumption is quite flexible and holds, at least approximately, in a sizable number of applications \cite{ahmed2014blind}. Through a novel application of the lifting trick, we show now it is possible to translate the measurement vector into a set of linear measurements with respect to the matrix $\bZ^{\star}=\bx\bh^T\in\mathbb{C}^{N\times L}$. 
While it is tempting to directly recover $\boldsymbol{Z}^{\star}$ from the obtained linear system of equations, it is under-determined since we have more unknowns, $NL$, than the number of observations, $N$. Fortunately, note that the columns of $\bZ^{\star}$ can be regarded as an ensemble of spectrally-sparse signals with the same spectral support, it is therefore possible to motivate this structure in the solution using the recently proposed atomic norm for spectrally-sparse ensembles \cite{chi2013joint,li2014off}. Specifically, we seek the matrix with minimum atomic norm that satisfies the set of linear measurements exactly in the noiseless setting, and within the noise level in the noisy setting. The proposed algorithm is referred to as {\em AtomicLift}. AtomicLift can be efficiently implemented via semidefinite programming using off-the-shelf solvers. Moreover, the spikes can be localized by identifying the peaks of a dual polynomial constructed from the dual solution of AtomicLift. Numerical examples are provided to demonstrate the effectiveness of AtomicLift in both noiseless and noisy settings.

To establish rigorous performance guarantees of AtomicLift in the noiseless case, we assume that each row of $\bB$ is identically and independently drawn from a distribution that obeys a simple isotropy property and an incoherence property, which is motivated by Cand\`es and Plan in their development of a RIPless theory of compressed sensing \cite{CandesPlan2011RIPless}. This implies the PSF to have certain ``spectral-flatness'' property, so that the PSF has on average the same energy at different frequencies. Moreover, this assumption is flexible to allow the entries in each row of $\bB$ to be correlated. On the other hand, we assume the minimum separation between spikes is at least $1/M$, where $N=4M+1$. This condition is the same as the requirement in \cite{candes2014towards,tang2012compressed} even when the PSF is known perfectly. Under these conditions, we show that in the noiseless setting, with high probability, AtomicLift recovers the spike signal model up to a scaling factor as soon as $N$ is on the order of $O(K^2L^2)$ up to logarithmic factors. Importantly, our result does not make randomness assumptions on the spike signal $\bx$ nor the orientation of the PSF in the subspace $\bh$. Recall that when the PSF is known exactly, it is capable to resolve $K$ spikes as soon as $N$ is on the order of $O(K)$. Therefore, when both $K$ and $L$ are not too large, AtomicLift is provably capable of blind spikes deconvolution at a price of more samples. The stability analysis of AtomicLift in the noisy setting is presented elsewhere \cite{chi2016robust} due to space limits.

Our proof is based on constructing a valid {\em vector-valued} dual polynomial that certifies the optimality of the proposed convex optimization algorithm with high probability. The construction is 
inspired by \cite{candes2014towards,tang2012compressed}, where the squared Fejer's kernel is an essential building block in the construction. Nonetheless, significant, and nontrivial, modifications are necessary since our dual polynomial is vector-valued rather than scalar-valued as in the existing works, and is additionally complicated by the special linear operator induced from lifting.

\subsection{Comparisons with Related Work}

 Our approach is inspired by the pioneering work of \cite{ahmed2014blind,ling2015self,candes2012phaselift}, which applied the lifting trick to quadratic and bilinear inverse problems such as phase retrieval and blind deconvolution. In \cite{ahmed2014blind}, both the PSF $\bg$ and the signal $\bx$ are assumed lying in some known subspaces with dimension $L$ and $K$ respectively. It is established in \cite{ahmed2014blind} that a nuclear norm minimization algorithm achieves exact recovery from a near-optimal number of samples $N\gtrsim O(K+L)$ up to logarithmic factors, as long as the subspace of $\bg$ is deterministic and satisfies certain spectral-flatness condition, and the subspace of $\bx$ is composed of i.i.d. Gaussian entries. Unfortunately, this algorithm cannot be applied in our setting, as the signal $\boldsymbol{x}$ does not lie in a {\em known} subspace, but rather an unknown subspace parameterized by the continuous-valued locations of the spikes.

In \cite{ling2015self}, Ling and Strohmer extended the framework in \cite{ahmed2014blind} to allow the signal $\boldsymbol{x}$ to be a $K$-sparse vector in a random Gaussian or random partial DFT matrix. It is established in \cite{ling2015self} that an $\ell_1$-minimization algorithm achieves exact recovery as soon as $N$ is on the order of $O(KL)$ up to logarithmic factors. If the locations of the spikes in $\boldsymbol{x}$ lies on the grid of the DFT frame, $\bx$ becomes a sparse vector in the DFT frame, it is possible to apply the algorithm proposed in \cite{ling2015self}. However, the performance guarantees in \cite{ling2015self} cannot be applied. Moreover, since the locations of the spikes do not necessarily lie on any a priori defined grid, it will encounter the basis mismatch issue discussed extensively in \cite{Chi2011sensitivity} that potentially results in significant performance degeneration. The same holds true for the algorithm of Lee et.~al. \cite{lee2015stability} which assumes $\bx$ is sparse in a pre-determined dictionary.

Finally, our work is related to recent advances in super-resolution \cite{candes2014towards,tang2012compressed,fernandez2015super}\nocite{bendory2015super,li2015stable}--\cite{duval2015exact} using total variation or atomic norm minimization, but significantly deviates from the existing literature since we focus on the more challenging case when the PSF is not known. To the best of the author's knowledge, our work provides the first algorithm for blind super-resolution with provable performance guarantees.

\subsection{Paper Organization}
The rest of the paper is organized as follows. Section~\ref{formulation} formulates the problem of blind spikes deconvolution, describes the proposed AtomicLift algorithm and its performance guarantees. Section~\ref{examples} provides numerical examples to demonstrate the performance of AtomicLift. Section~\ref{proof_main_theorem} proves the main theorem in this paper. Finally, we conclude and outline a few future directions in Section~\ref{conclusions}.
   
Throughout the paper, we use boldface capital letters to denote matrices $\bA$ and vectors $\ba$, $\left(\cdot\right)^{T}$ to denote the transpose, $\left(\cdot\right)^{H}$ to denote the conjugate transpose, and $\left(\cdot\right)^*$ to denote the conjugate. $\left\|\bA\right\|_{\mathrm{F}}$, $\left\|\bA\right\|$ denote the Frobenius norm and the spectral norm of a matrix $\bA$ respectively, and $\left\|\ba\right\|_2$ denotes the $\ell_2$ norm of a vector $\ba$.

\section{AtomicLift for Blind Sparse Spikes Deconvolution}\label{formulation}
We will first describe the problem formulation and the AtomicLift algorithm in the noiseless case, and then extends to the noisy case.
\subsection{Problem Formulation}
Let $x(t)$ be a continuous-time spike signal given as
$$x(t) = \sum_{k=1}^{K}\bar{a}_{k}\delta(t-\bar{\tau}_{k}),$$
where $K$ is the number of spikes, $\bar{a}_k\in\mathbb{C}$ and $\bar{\tau}_k\in [0,T_{\max})$ are the complex amplitude and delay of the $k$th spike, $1\leq k\leq K$, and $T_{\max}$ is the maximum allowable delay of the spikes. Let $g(t)$ be the PSF with the bandwidth $[-B_{\max},B_{\max}]$. The convolution of $g(t)$ and $x(t)$ is given as \begin{align}\label{sparse_con}
y(t) &=   x( t )* g(t)   =  \sum_{k=1}^{K} \bar{a}_{k}g(t- \bar{\tau}_{k}) ,  
\end{align}
where $*$ denotes convolution. Taking the Fourier transform of \eqref{sparse_con}, we have
\begin{equation}\label{receive_freq}
Y(f) = X(f)G(f) = \left( \sum_{k=1}^{K}\bar{a}_{k}e^{-j2\pi f \bar{\tau}_{k}}  \right) \cdot G(f)
\end{equation}
for $f\in  [-B_{\max},B_{\max}]$, where $X(f)$, $G(f)$, and $Y(f)$ are the Fourier transforms of $x(t)$, $g(t)$, and $y(t)$ respectively. In order to digitally process the output, we will uniformly sample \eqref{receive_freq} at $N=4M+1$ points $f_n = \frac{B_{\max}n}{2M}$, $n = -2M, \ldots, 2M$, and denote $x_n \triangleq X(f_n)$, $g_n \triangleq G(f_n)$, and $y_n  \triangleq Y(f_n) $. This yields 
\begin{align}
y_n   &= x_n \cdot g_n =  \left( \sum_{k=1}^{K}\bar{a}_{k}e^{-j2\pi n{\tau}_{k} \frac{B_{\max}T_{\max}}{2M} }  \right) \cdot g_n ,  \label{unnormalized_freq}
\end{align}
where $\tau_k = \bar{\tau}_k/T_{\max}\in [0,1)$ is the normalized delay. From \eqref{unnormalized_freq}, it is straightforward to see that the number of samples needs to satisfy $2M\geq B_{\max}T_{\max}$ so that the delays $\tau_{k} \in [0,1)$ can be uniquely identified. Since we're interested in algorithmic frameworks that allow identification of the spike signal using as small $M$ as possible, without loss of generality, we will assume $2M = B_{\max}T_{\max}$, and consider the normalized delays $\tau_k\in [0,1)$ in this paper, and the sample complexity $2M$ becomes the bandwidth of $g(t)$.

We can now rewrite \eqref{unnormalized_freq} as
\begin{equation}\label{freq}
y_{n} = g_n \cdot x_n  =  g_{n}\cdot\left(\sum_{k=1}^{K} \bar{a}_{k}e^{-j2\pi n\tau_{k}} \right),
\end{equation}
where $x_n =\sum_{k=1}^{K}\bar{a}_{k}e^{-j2\pi n\tau_{k}}$, for $n = -2M, \ldots, 2M$.  Denote the set of spike locations as $\mathcal{T} = \{\tau_k\}_{k=1}^K$. In the matrix form, we rewrite \eqref{freq} as
\begin{equation} \label{freq_vec}
\y = \mbox{diag}(\bg) \bx,
\end{equation}
where $\y = [y_{-2M}, \ldots, y_{2M}]^T$, $\bx = [x_{-2M}, \ldots, x_{2M}]^T $, and $\bg = [g_{-2M}, \ldots, g_{2M}]^T $. Interestingly, \eqref{freq_vec} is also related to blind calibration for uniform linear arrays, where $\bg$ can be interpreted as the vector of array antenna gains. 

Clearly, there is an unavoidable scaling ambiguity for the identification of $\bg$ and $\bx$ from $\y$, since for any nonzero scalar $\beta$, $\by = \mbox{diag}(\beta \bg)\bx =  \mbox{diag}(\bg)(\beta\bx)$. Our goal is to recover both $\boldsymbol{g}$ and $\bx$, in particular, the set of spikes $\mathcal{T}$ with their corresponding amplitudes up to a scaling factor.

\subsection{AtomicLift} \label{algorithm}
The problem of blind spikes deconvolution is extremely ill-posed without further constraints \cite{li2015identifiability}. In this paper, inspired by \cite{ahmed2014blind}, we make the assumption that $\bg$ lies in a known low-dimensional subspace, given as 
$$\bg = \bB\bh,$$ 
where $\bB\in\mathbb{C}^{N\times L}$ is known, $\bh\in\mathbb{C}^{L}$ is unknown, and $L\ll N$. Denote $\bB^T=[\bb_{-2M}, \ldots, \bb_{2M}]$, where $\bb_n \in \mathbb{C}^{L}$ is the $n$th column of the matrix $\bB^T$. We rewrite $y_n$ in \eqref{freq} as
\begin{equation}\label{lifting}
y_n  = \boldsymbol{b}_n^T \boldsymbol{h} x_n =  \boldsymbol{b}_n^T \boldsymbol{h}   \boldsymbol{e}_n^T \boldsymbol{x}   =  \be_n^T (\bx \bh^T) \bb_n,  
\end{equation}
where $\boldsymbol{e}_n$ is the $n$th standard basis vector of $\mathbb{R}^N$. Let $\boldsymbol{Z}^{\star} = \bx \bh^T$, using the lifting trick \cite{ahmed2014blind,ling2015self,candes2012phaselift}, \eqref{lifting} can be rewritten as 
a linear measurement of $\bZ^{\star}$, 
$$y_n =   \be_n^T \bZ^{\star} \bb_n = \langle \bZ^{\star}, \be_n\bb_n^H \rangle, \quad n=-2M,\ldots, 2M,$$ 
where $ \langle \bY,\bX \rangle= \trace(\bX^H\bY)$.
Therefore, $\y$ can be regarded a set of linear measurements of $\bZ^{\star}$, i.e. 
\begin{equation}\label{measurements}
\y = \mathcal{X}(\bZ^{\star}),
\end{equation} 
where $\mathcal{X}:\mathbb{C}^{N\times L}\mapsto \mathbb{C}^{N}$ denotes the operator that performs the linear mapping \eqref{lifting}. 

Since from $\bZ^{\star}$, we can recover $\bx$ and $\bg$, up to a scaling factor, as the left and right singular vectors of ${\bZ}^{\star}$, respectively, we now wish to recover the matrix $\bZ^{\star}$ from $\y$. Though it appears the number of  unknowns is much more than the number of measurements, a key observation is that $\bZ^{\star}$ can be regarded as a signal ensemble where each column signal is composed of $K$ complex sinusoids with the same frequencies, 
$$\boldsymbol{Z}^{\star} = \boldsymbol{x}\boldsymbol{h}^T =  \sum_{k=1}^K a_k \bc(\tau_k)\bh^T \in \mathbb{C}^{N \times L},$$
where $\ba =\sqrt{N}[\bar{a}_1,\ldots,\bar{a}_K]$, and
$$\bc(\tau)  = \frac{1}{\sqrt{N}}\left[e^{-j2\pi(-2M)\tau}, \ldots, 1,\ldots, e^{ -j2\pi (2M)\tau}\right]^T $$
represents a complex sinusoid with the frequency $\tau\in[0,1)$. Therefore, it is possible to motivate the joint spectral sparsity of the columns of $\bZ^{\star}$ by minimizing the atomic norm \cite{chandrasekaran2012convex} for joint spectrally-sparse ensembles proposed by the author in \cite{chi2013joint}. To proceed, define the set of atoms as 
$$\cA=\left\{\bA(\tau,\bu)=\bc(\tau)\bu^H  \in\mathbb{C}^{ N \times L} | \tau \in [0,1), \|\bu \|_2 =1 \right\}.$$ 
The atomic norm seeks the tightest convex relaxation of decomposing a matrix $\bZ\in \mathbb{C}^{N\times L}$ into the smallest number of atoms in $\mathcal{A}$, and is defined as \cite{chi2013joint}
\begin{align}
& \| \bZ\|_{\cA} = \inf \left\{ t>0: \; \bZ\in t\; \mbox{conv}(\cA) \right\}\label{atomic_def}  \\
&= \inf_{\stackrel{\tau_k\in [0,1)}{\bu_k\in\mathbb{C}^L: \|\bu_k\|_2=1}} \left\{ \sum_k c_k  \Big| \bZ = \sum_k c_k \bA(\tau_k,\bu_k), c_k \geq 0 \right\}, \nonumber
\end{align}
where $\mbox{conv}(\cA)$ is the convex hull of $\cA$. The corresponding decomposition of $\bZ$ that achieves the atomic norm is called the {\em atomic decomposition}. Moreover, $\| \bZ\|_{\cA}$ admits an equivalent semidefinite programming (SDP) characterization \cite{chi2013joint} which can be computed efficiently using off-the-shelf solvers:
\begin{align*}
 \| \bZ\|_\cA = \inf_{\bu\in\mathbb{C}^N,\bW\in\mathbb{C}^{L\times L}}  \Big\{ & \frac{1}{2}\trace(\toep(\bu)) + \frac{1}{2}\trace(\bW) \Big| \\
 &  \begin{bmatrix}
\toep(\bu) & \bZ \\
\bZ^H & \bW \end{bmatrix} \succeq \bf{0} \Big\},
\end{align*} 
where $\toep(\bu)$ is the Toeplitz matrix with $\bu$ as the first column. We then propose the following algorithm, denoted as  {\em AtomicLift}, to motivate the joint spectral sparsity of $\boldsymbol{Z}$ by seeking the matrix with the smallest atomic norm satisfying the measurements:
\begin{equation}\label{atomic_lifting}
\hat{\bZ}=  \argmin_{\bZ\in\mathbb{C}^{N\times L}} \; \| \boldsymbol{Z} \|_{\mathcal{A}} \quad \mbox{s. t.} \quad \y = \mathcal{X}(\bZ).
\end{equation}

\subsection{Performance Guarantee of AtomicLift}

The main result of this paper is that if we assume the rows of the subspace $\bB$ are i.i.d. drawn from some distribution that satisfies the isotropy property and the incoherence property, together with a mild separation condition for the spike signal, the proposed AtomicLift algorithm provably recovers the PSF as well as the spike signal, up to a scaling ambiguity, with high probability as long as $M$ is large enough.

Specifically, we assume each row of $\bB$ is sampled independently and identically from a population $F$, i.e. $\bb_n\sim F$, $n=-2M,\ldots, 2M$. Furthermore, we require $F$ satisfies the following properties:
\begin{itemize}
\item {\em \yc{Isotropy} property:} $F$ is said to satisfy the \yc{isotropy} property if for $\bb\sim F$, 
$$\mathbb{E}\bb\bb^H = \bI_L.$$
\item {\em Incoherence property:} for $\bb = [b_1, \ldots, b_L]^T\sim F$, define the coherence parameter $\mu$ of $F$ as the smallest number that 
$$ \max_{1\leq i \leq L} \; |b_i|^2 \leq \mu$$ holds.
\end{itemize}
The above definitions are motivated by \cite{CandesPlan2011RIPless} in the development of a RIPless theory of compressed sensing. In particular, the incoherence parameter $\mu$ is a deterministic bound on the maximum entry of $\bb$, which can be extended to a stochastic setting using the stochastic incoherence discussed in \cite{CandesPlan2011RIPless}, so that $F$ is allowed to be composed of unbounded sub-Gaussian or sub-exponential distributions. It is possible to extend our results to the stochastic setting in \cite{CandesPlan2011RIPless}, but for conciseness in this paper, we'll limit ourselves to the deterministic setting. 

We discuss the implications of the above properties for  the PSF $\bg=\bB\bh$ when $\bh$ is arbitrary. Following the isotropy property, we have $\mathbb{E}|g_n|^2 =\| \bh\|_2^2$ for all $n$, which means that on average, the PSF is ``spectrally flat'', having the same energy across different frequencies. Also, $\mu\geq 1$ following the isotropy property, where the lower bound can be met, for example by selecting $\bb\sim F$ to be in the form of 
\begin{equation}\label{eg_mu}
\bb = [1, e^{j2\pi f}, \ldots, e^{j2\pi(L-1)f}],
\end{equation}
where $f$ is chosen uniformly at random in $[0,1]$.

Furthermore, define the minimum separation of the spike signal as 
 $$\Delta=\min_{1\leq i< j\leq K}|\tau_i -\tau_j|,$$ 
which is evaluated as the wrap-around distance on the unit circle. The performance guarantee of AtomicLift is presented in Theorem~\ref{main_theorem}, which is proven in Section~\ref{proof_main_theorem}.
\begin{theorem}\label{main_theorem}
Let $M\geq 64$. Assume $\bg$ lies in a random subspace $\bB$ whose rows are sampled i.i.d. from a population $F$ satisfying the isotropy property and the incoherence property, with the coherence parameter $\mu$. If $\Delta \geq 1/M$, then there exists a numerical constant $C$ such that 
$$ M \geq C   \mu K^2\yc{L^2} \log^2\left(\frac{M}{\delta}\right)$$
is sufficient to guarantee that we can recover $\bZ^{\star}=\bx\bh^T$ via the AtomicLift algorithm with probability at least $1-\delta$.
\end{theorem}
Theorem~\ref{main_theorem} allows $\bg$ to have arbitrary orientation in the subspace $\bB$, and applies to any {\em deterministic} spike signal as long as it satisfies the separation condition $\Delta\geq 1/M$ regardless of the amplitudes. The separation is the same as required by Cand\`es and Fernandez-Granda \cite{candes2014towards} for spikes deconvolution using total variation minimization even when the PSF is known perfectly. Therein they established that $N=O(K)$ measurements are sufficient to exactly recover the spike signal. In comparison, our performance guarantee is probabilistic that holds with high probability. 

Theorem~\ref{main_theorem} suggests that as long as $M$ is on the order of $O(K^2L^2)$ up to logarithmic factors, AtomicLift provably recovers the spike signal with high probability, as long as the conditions in Theorem~\ref{main_theorem} are satisfied. If $L$ is a small constant independent of $K$ and $M$, our bound simplifies to $M/\log^2 M\gtrsim O(K^2 )$, which suggests blind spike deconvolution is possible at a cost of more measurements. 

\subsection{AtomicLift for noisy data}\label{noisy_extension}
We consider a noisy version of \eqref{freq}, where the frequency-domain data samples are contaminated by additive noise: 
\begin{equation}\label{freq_noisy}
y_n = x_n \cdot g_n + w_n,
\end{equation}
where $\bw=[w_{-2M},\ldots, w_{2M}]^T$ is bounded by $\|\bw\|_2\leq \epsilon$. Accordingly, the lifted measurement model becomes
\begin{equation}\label{measurements_noisy}
\y = \mathcal{X}(\bZ^{\star}) + \bw.
\end{equation}
The AtomicLift algorithm can be modified with the measurement constraint that obeys the noise level, given as
\begin{equation}\label{atomic_lifting_noisy}
\hat{\bZ}_{\text{noisy}}=  \argmin_{\bZ\in\mathbb{C}^{N\times L}} \; \| \boldsymbol{Z} \|_{\mathcal{A}} \quad\mbox{s.t.}\quad  \left\| \y -\mathcal{X}(\bZ) \right\|_2\leq \epsilon. 
\end{equation}

\begin{figure*}[ht]
\begin{center}
\begin{tabular}{cccc}
\hspace{-0.2in}\includegraphics[width=0.26\textwidth]{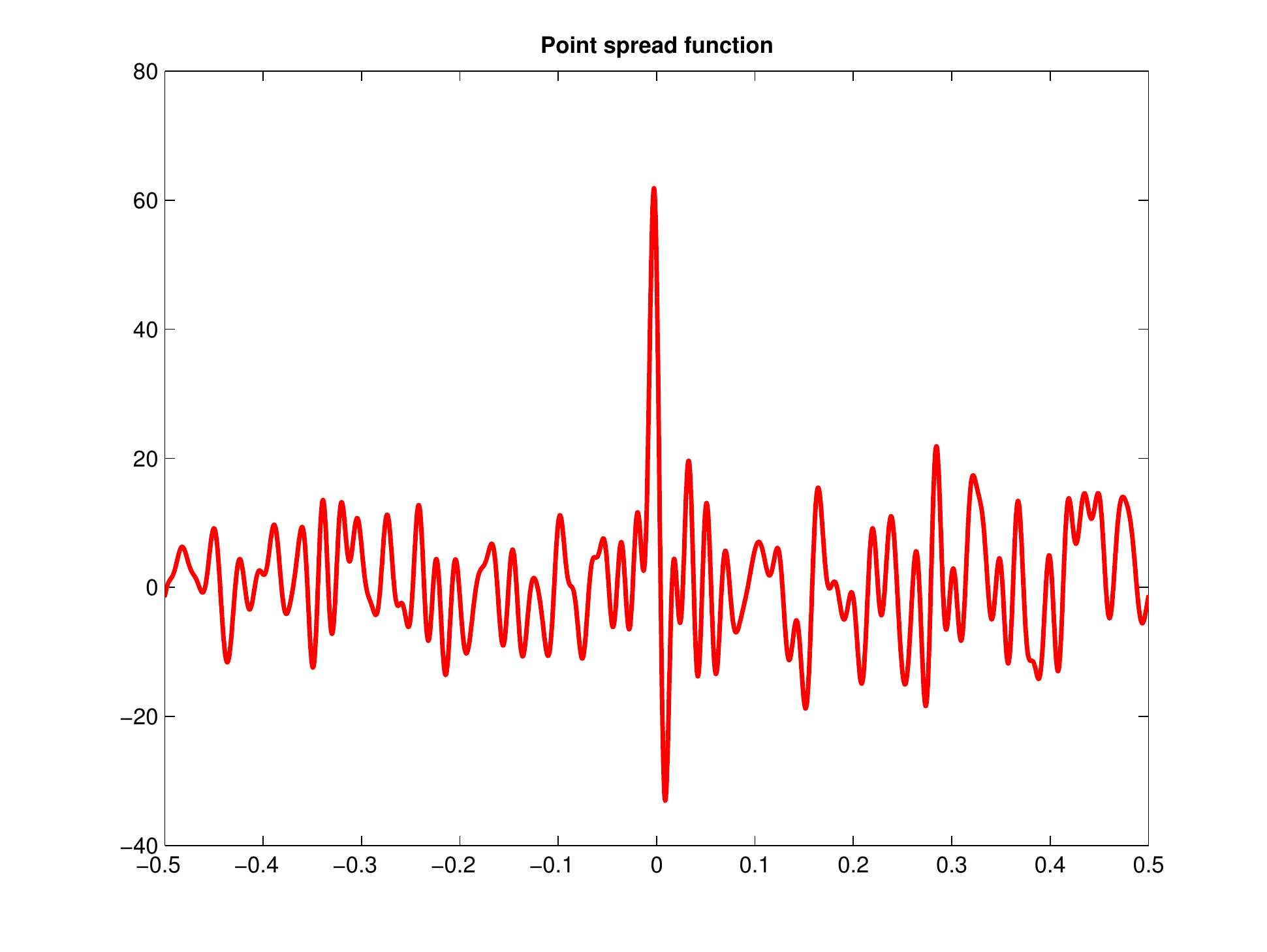} & 
\hspace{-0.2in}\includegraphics[width=0.26\textwidth]{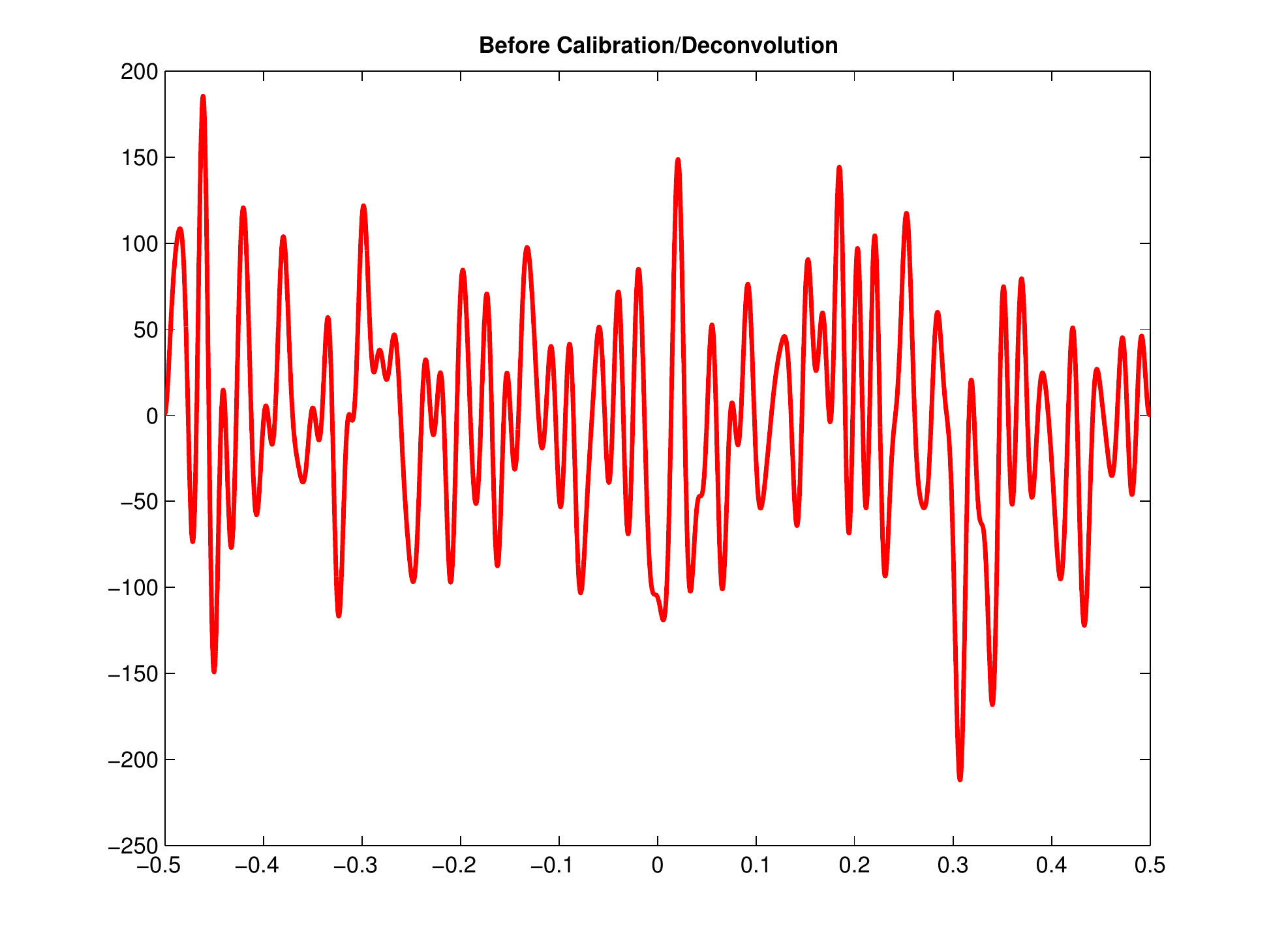} &
\hspace{-0.2in}\includegraphics[width=0.26\textwidth]{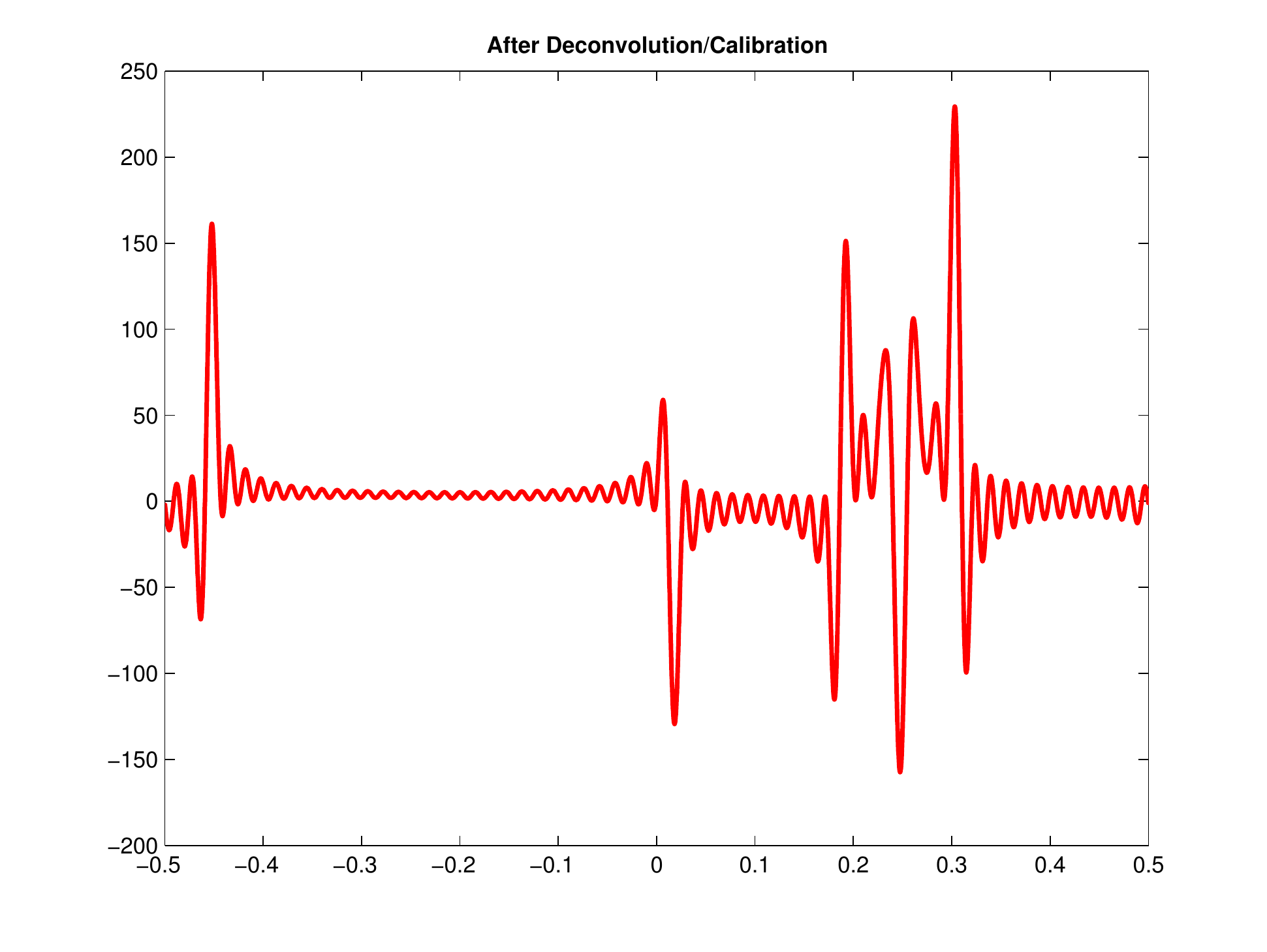}&\hspace{-0.2in}\includegraphics[width=0.26\textwidth]{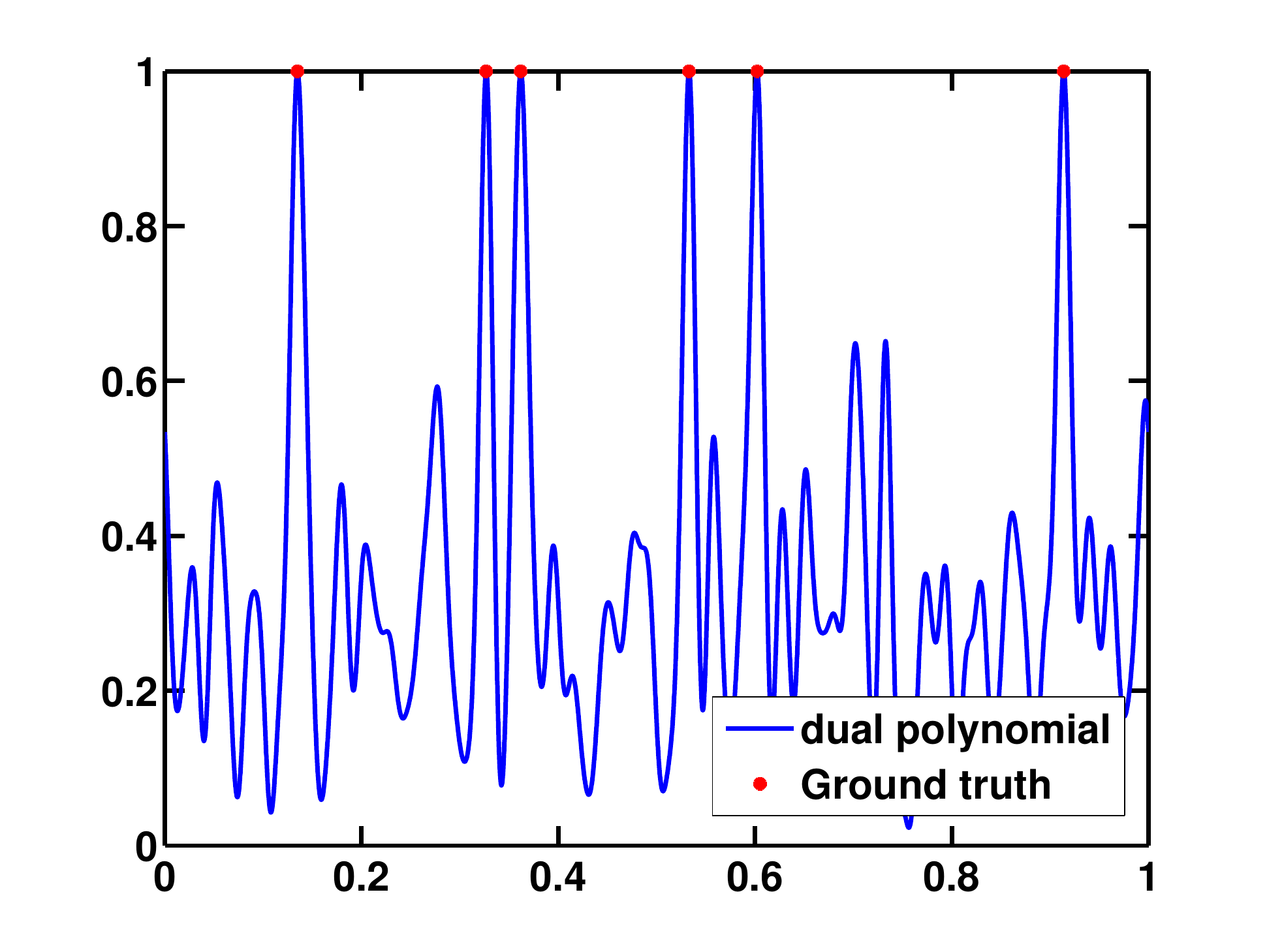} \\
(a) PSF & \hspace{-0.2in}(b) Convolution with the PSF &\hspace{-0.2in} (c) Deconvolution &\hspace{-0.2in} (d) Localization  
\end{tabular}
\end{center}
\caption{Blind spikes deconvolution using AtomicLift: (a) PSF; (b) convolution between the PSF in (a) and a sparse spike signal; (c) deconvolution with the PSF using (b); \yc{(d) exact localization of the spikes via the dual polynomial}.}\label{example}
\end{figure*} 

\subsection{Spike Localization}
  
Define $\langle \bY,\bX \rangle_{\mathbb{R}}=\mbox{Re}(\langle \bY,\bX \rangle)$. The dual norm of $\|\cdot\|_\cA$ can be defined as \cite{chi2013joint}
\begin{align*}
\|\bY\|_{\cA}^{\star} & = \sup_{\tau \in[0,1)} \left\| \bY^H \bc(\tau)  \right\|_2 .
\end{align*}

The dual problem of \eqref{atomic_lifting} can thus be written as
\begin{equation} \label{dual_lifting}
\hat{\bp} = \argmax_{\bp\in\mathbb{C}^N} \; \langle \bp , \y \rangle_{\mathbb{R}} \quad \mbox{s.t.} \quad \|\mathcal{X}^*(\bp)\|_{\cA}^{\star} \leq 1,
\end{equation}
and the dual problem of \eqref{atomic_lifting_noisy} can be written as
\begin{equation} \label{dual_lifting_noisy} 
\hat{\bp} = \argmax_{\bp\in\mathbb{C}^{N}} \langle \bp, \y\rangle -\frac{\epsilon}{2} \| \bp\|_2, \quad \mbox{s.t.} \quad \|\mathcal{X}^*(\bp)\|_{\mathcal{A}}^* \leq 1,
\end{equation}
where $\mathcal{X}^*(\bp) =\sum_{n=-2M}^{2M} p_n \be_n\bb_n^H$. Write the vector-valued dual polynomial $\bQ(\tau)\in\mathbb{C}^{L}$ as
\begin{align*} 
\bQ(\tau) &=( \mathcal{X}^*(\hat{\bp}) )^H  \bc(\tau),
\end{align*}
where $\hat{\bp}$ is the solution of the dual problems \eqref{dual_lifting} and \eqref{dual_lifting_noisy}, then the spikes can be localized by the peaks of $ \left\| \bQ(\tau) \right\|_2$:
\begin{equation}\label{localization}
\hat{\mathcal{T}} = \left\{ \tau\in [0,1)  | \left\| \bQ(\tau) \right\|_2 = 1 \right\}.
\end{equation} 
\yc{We refer the readers to standard arguments in \cite{tang2012compressed,chi2datomic,li2014off} for more details.}

 \begin{figure*}[ht]
\begin{center}
\begin{tabular}{cccc}
\hspace{-0.1in}\includegraphics[width=0.25\textwidth]{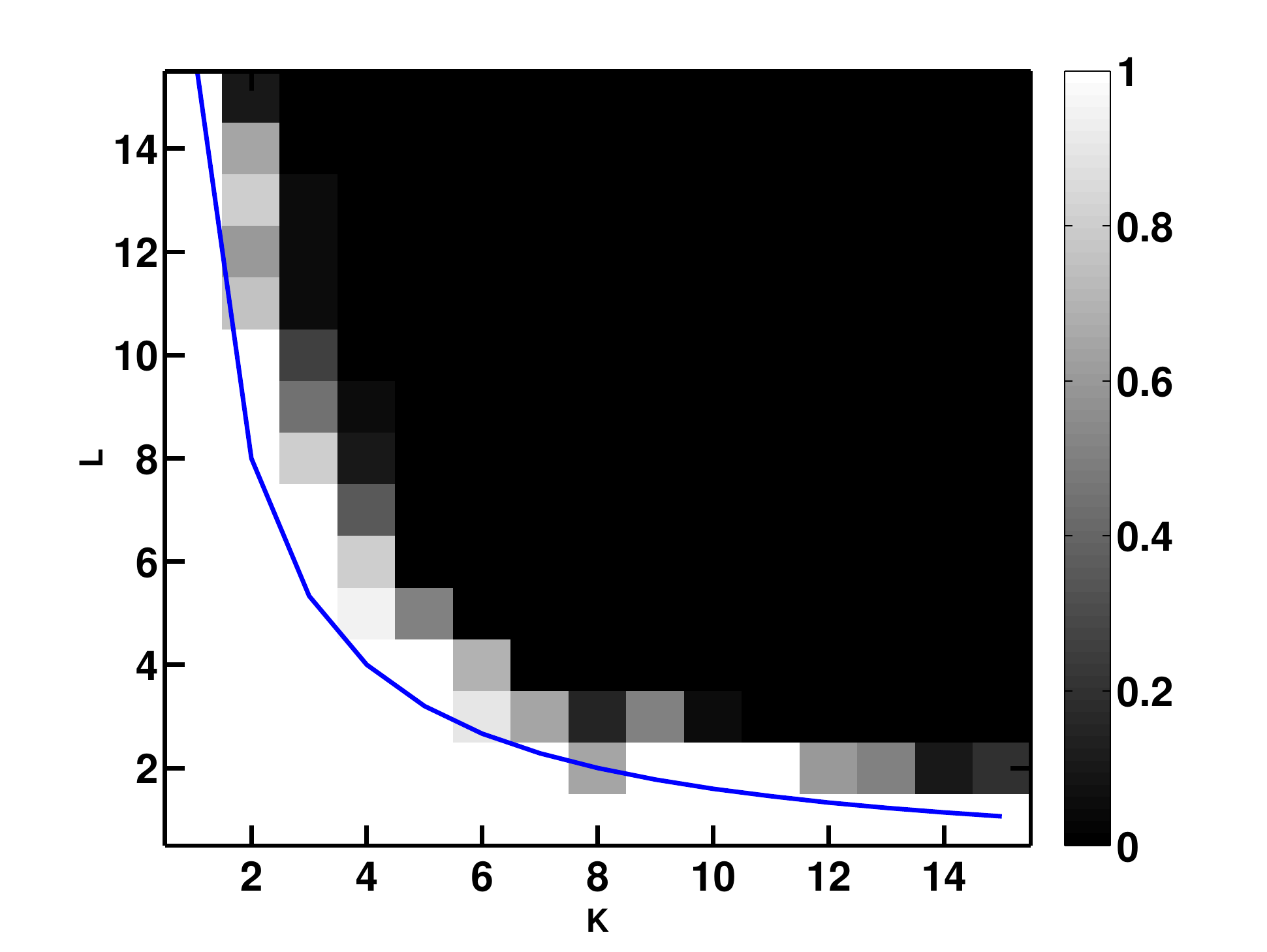} &\hspace{-0.1in}\includegraphics[width=0.25\textwidth]{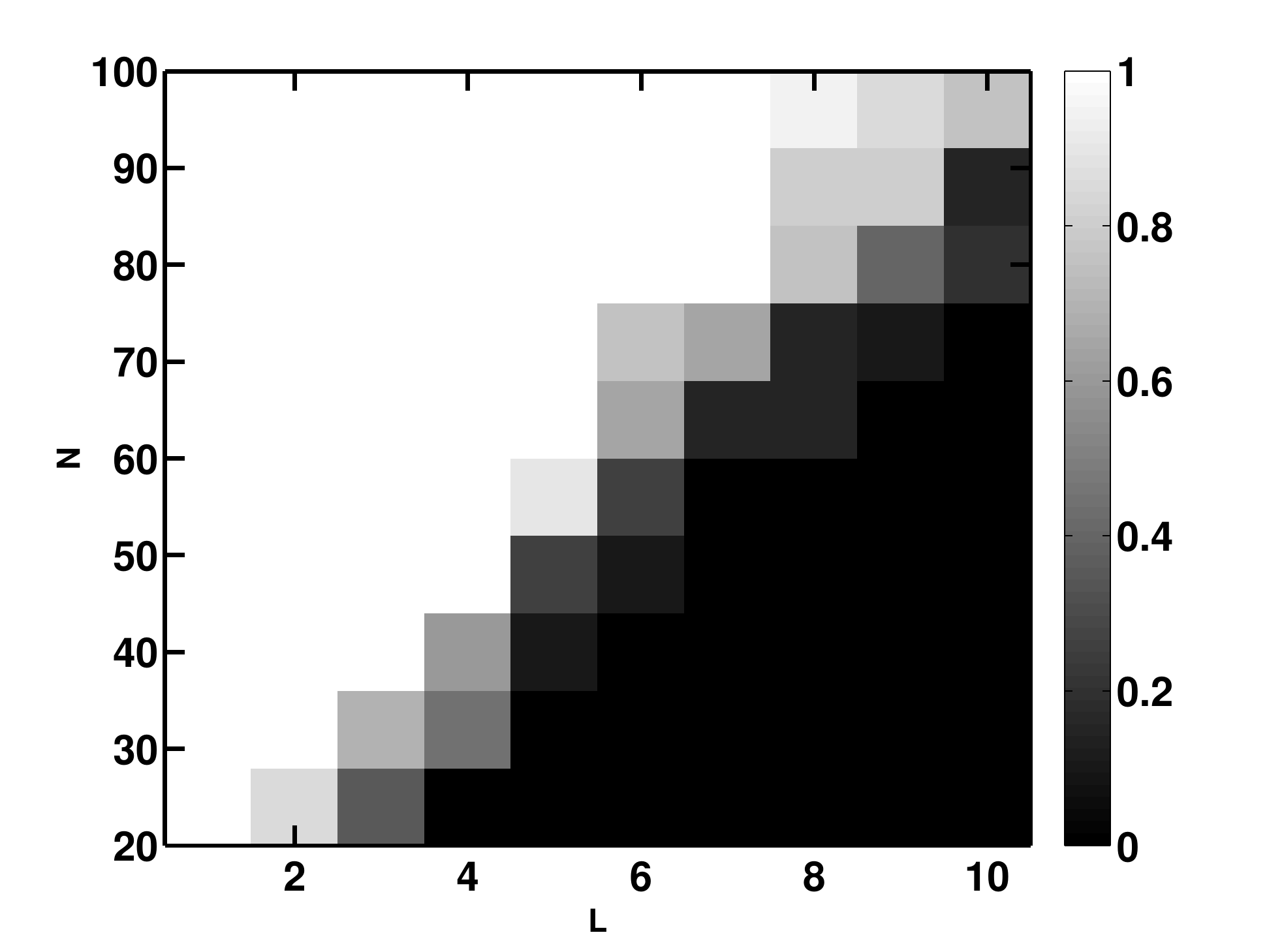} &
\hspace{-0.1in}\includegraphics[width=0.25\textwidth]{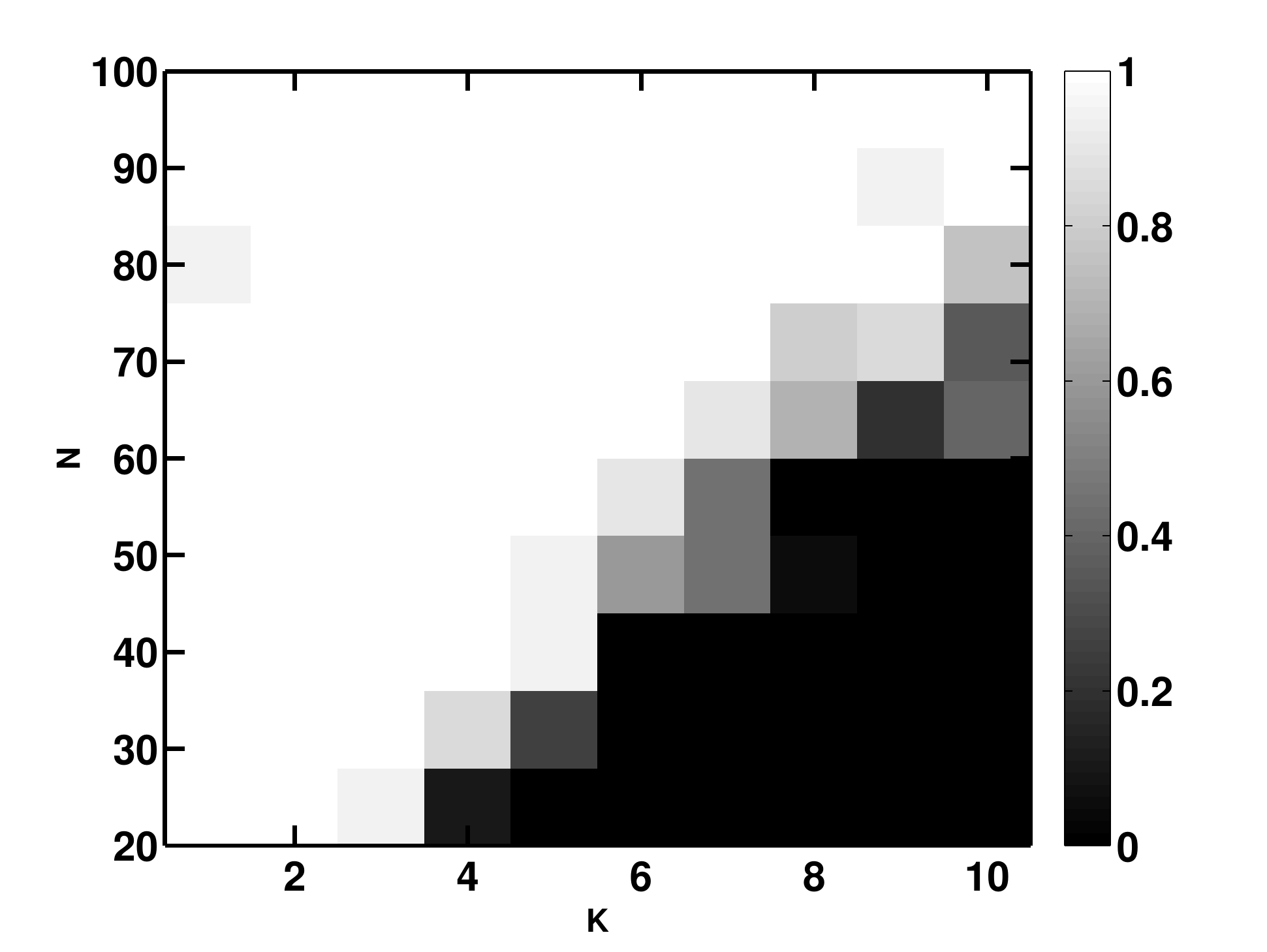} &
\hspace{-0.1in}\includegraphics[width=0.25\textwidth]{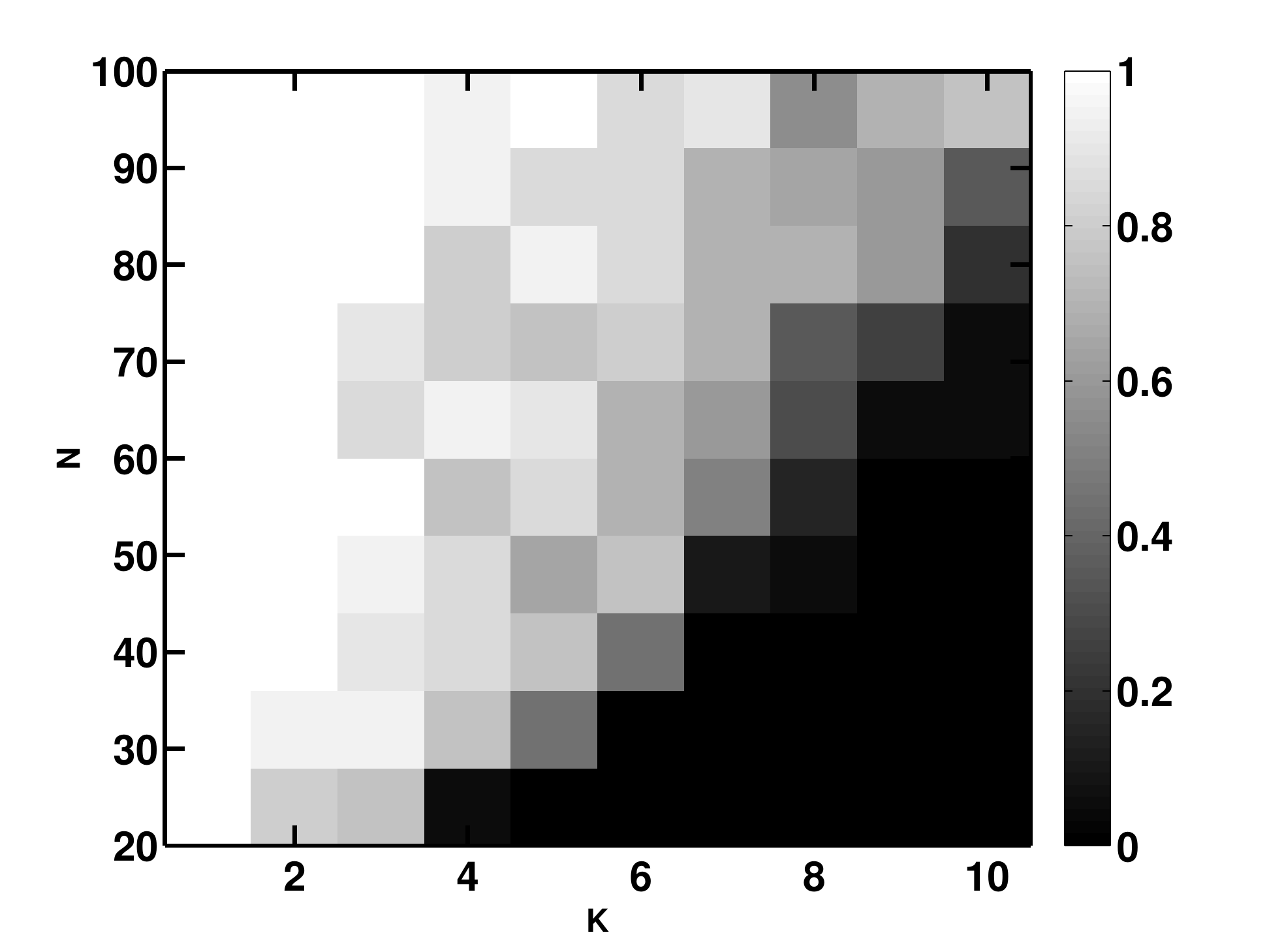}  \\
\hspace{-0.2in}(a) & \hspace{-0.2in}(b)& \hspace{-0.2in}(c)& \hspace{-0.2in}(d)
\end{tabular}
\end{center}
\caption{The average success rate of AtomicLift (a) with respect to the number of spikes $K$ and the subspace dimension $L$ when the number of measurements $N = 64$; (b) with respect to $K$ and  $N$ when $L=3$; (c) with respect to $L$ and $N$ when $K=4$ for spikes generated satisfying a separation condition $\Delta\geq 1/N$; (d) with respect to $L$ and $N$ when $K=4$ for randomly generated spike locations.}\label{fixNKL}
\end{figure*}

\section{Numerical Experiments} \label{examples}
We perform a series of numerical experiments to validate the performance of AtomicLift implemented using MOSEK \cite{grant2008cvx}\footnote{The code can be downloaded from \url{http://www2.ece.ohio-state.edu/~chi/papers/atomiclift.m}.}. Without loss of generality, in all numerical experiments, we set the index $n\in\{0,\ldots, N-1\}$, rather than $n\in\{-2M,\ldots, 2M\}$ as in the previous sections.

Let $N =64$. We first generate the spike locations uniformly at random, respecting the minimum separation $\Delta \geq 1/N$ (which is smaller than what the theory requires), whose coefficients are generated with a dynamic range of $10$dB and uniform phase. We generate the subspace $\bB$ by selecting each row i.i.d. following \eqref{eg_mu} with $L=3$, and choose the coefficient vector $\bh$ as an all-one vector. Fig.~\ref{example}~(a) shows the PSF in the time domain, and the convolution with a spike signal with $K=6$ spikes is shown in (b). If the PSF is known, one can deconvolve (b) with the PSF and obtain the calibrated time-domain signal in (c). Clearly Fig.~\ref{example}~(c) is very different from Fig.~\ref{example}~(b), therefore calibration must be performed if the PSF is unknown to avoid severe performance degeneration. Fig.~\ref{example} (d) demonstrates the exact localization of the spikes via computing the dual polynomial of AtomicLift.

We next examine phase transition of the AtomicLift algorithm. We randomly generate the low-dimensional subspace $\bB$ with i.i.d. standard Gaussian entries, and the coefficient vector $\bh$ with i.i.d. standard Gaussian entries. The spike signal is generated in the same fashion as above. For each simulation, we compute the normalized error $\|\hat{\bZ} - \bZ^{\star} \|_{\mathrm{F}} / \| \bZ^{\star}\|_{\mathrm{F}}$, where $\hat{\bZ}$ is the estimate of the lifted matrix $\bZ^{\star} = \bx \boldsymbol{h}^T$. First fix $N=64$. For each pair of $(K,L)$, we run $20$ Monte Carlo simulations of the AtomicLift algorithm, and claim the reconstruction of a simulation is successful if the normalized error is below $10^{-3}$. Fig.~\ref{fixNKL}~(a) shows the average success rate with respect to the number of spikes $K$ and the subspace dimension $L$. For comparison, we also plot the hyperbola curve $KL=20$ which roughly matches the phase transition boundary. Similarly, Fig.~\ref{fixNKL} (b) shows the average success rate with respect to $L$ and $N$ for a fixed $K = 4$, and Fig.~\ref{fixNKL} (c) shows the average success rate with respect to $K$ and $N$ for a fixed $L = 3$. The phase transition plots suggest that AtomicLift succeeds when $N\gtrsim O(KL)$, which is better than the prediction of our theory. Fig.~\ref{fixNKL} (d) shows the average success rate with respect to $K$ and $N$ as in the same setting of Fig.~\ref{fixNKL} (b), except that the locations of the spikes are randomly generated without obeying the separation condition. It can be seen that the phase transition is not as sharp, which is \yc{in line with} existing results in applying atomic norm minimization to spectrum estimation \cite{tang2012compressed,chi2datomic,li2014off}.

\begin{figure}[htp]
\begin{center}
\begin{tabular}{cc}
\hspace{-0.1in}\includegraphics[width=0.25\textwidth]{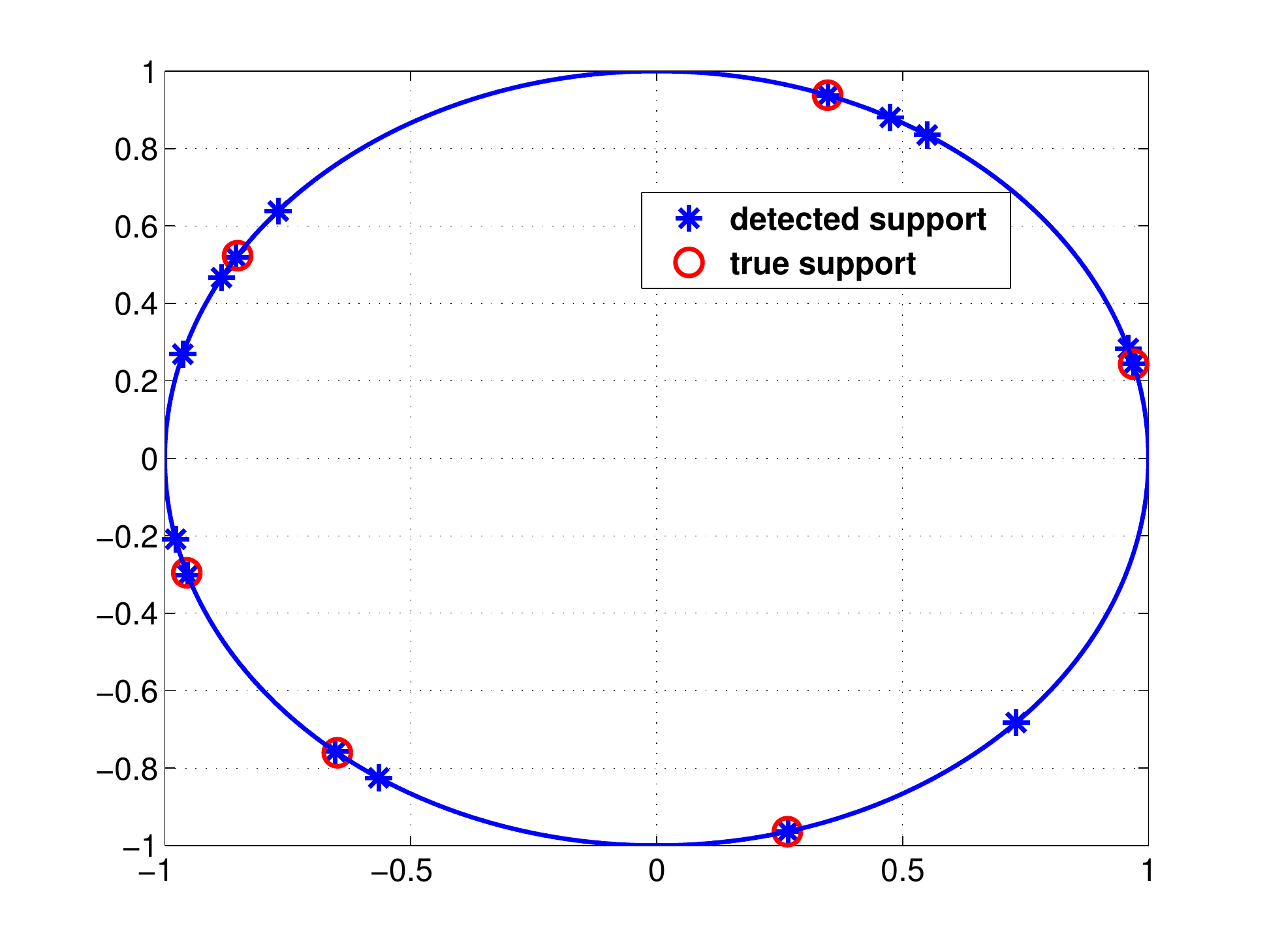} &
\hspace{-0.1in}\includegraphics[width=0.25\textwidth]{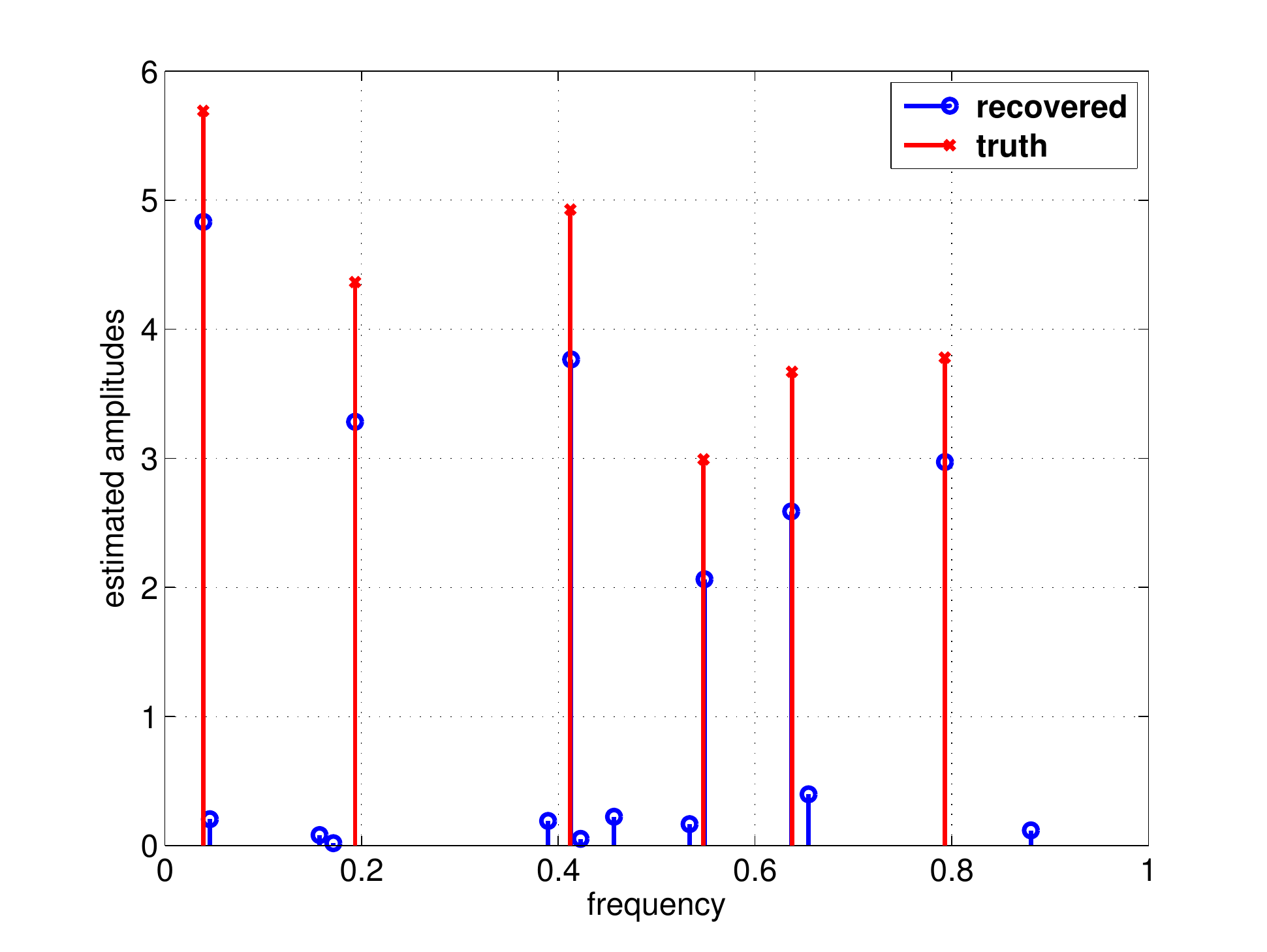} \\
(a) & (b)  
\end{tabular}
\end{center}
\caption{Source localization of AtomicLift in the noisy setting. (a) and (b): source locations identified using the dual polynomial and the recovered magnitudes, when $K=6$, $N=64$, $L=3$ and $\textrm{SNR}=15$dB.}\label{fig:noisy_localization}
\end{figure}

Finally, we examine the performance of AtomicLift in the noisy setting. Using similar setup as Fig.~\ref{fixNKL} when $N=64$ and $L=3$, we introduce additive white Gaussian noise as in \eqref{freq_noisy}, where each $w_n$ is i.i.d. generated with $\mathcal{CN}(0,\sigma^2)$. The signal-to-noise ratio (SNR) is defined as $10\log_{10}(\| \mathcal{X}(\bZ^{\star})\|_2^2/(N\sigma^2))$dB. Using a standard tail bound $\mathbb{P}\left(\|\bw\|_2\leq \sigma\sqrt{N+\sqrt{2N\log 2N}}\right)\geq 1-(2N)^{-1}$ \cite{CaiWang}, we set $\epsilon: = \sigma\sqrt{N+2\sqrt{N\log N}}$ in \eqref{atomic_lifting_noisy}. We use \eqref{localization} to identify the spike locations. Fig.~\ref{fig:noisy_localization} shows the recovered source locations and their magnitudes. It is worth noting that the dual polynomial will {\em overestimate} the number of sources, and further model order estimation is still necessary for noisy data. 

\section{Proof of Main Theorem}\label{proof_main_theorem}
In this section we prove Theorem~\ref{main_theorem}. \yc{First, we describe the desired form of a valid vector-valued dual polynomial and its properties that will guarantee the optimality of AtomicLift in Proposition~\ref{prop_optimality}. We next design the dual polynomial with the help of the squared Fejer's kernel \cite{candes2014towards} in Section~\ref{sec:construction_dual}. The rest of the proof is then to carefully validate it satisfies all the required properties.} Proposition~\ref{prop_optimality} presents a sufficient condition for the optimality of AtomicLift in \eqref{atomic_lifting}, whose proof is provided in Appendix~\ref{proof_optimality}.
\begin{prop}\label{prop_optimality}
The solution to \eqref{atomic_lifting} is unique if there exists a vector $\bq \in \mathbb{C}^N$ such that the vector-valued dual polynomial 
\begin{align}\label{form_of_Q}
\bQ(\tau) &=( \mathcal{X}^*(\bq) )^H  \bc(\tau) = \bB^T\diag(\bq) \bc(\tau) \nonumber \\
&= \frac{1}{\sqrt{N}}\sum_{n=-2M}^{2M}  e^{-j2\pi \tau n} q_n^* \bb_n \in\mathbb{C}^L
\end{align} 
satisfies 
\begin{subequations}
\begin{align}
& {\bQ}(\tau_k)     = \frac{1}{\|\bh\|_2^2}\mbox{sign}(a_k^* ) \bh^* \quad  \forall \tau_k \in \mathcal{T}, \label{condition_one} \\
 & \left\|\bQ(\tau) \right\|_2  < 1, \quad \forall \tau\in [0,1]\backslash \mathcal{T}, \label{condition_two}
\end{align}
\end{subequations}
where $\mbox{sign}(\cdot)$ is the complex sign function.
\end{prop}
Proposition~\ref{prop_optimality} suggests that the solution to AtomicLift in \eqref{atomic_lifting} is exact and equals to $\bZ^{\star}$ if we can find a dual polynomial $\bQ(\tau)$ with the valid form that satisfies \eqref{condition_one} and \eqref{condition_two}. Our objective is then to construct such a valid dual polynomial.

\newcounter{tempequationcounter}
\begin{figure*}[htb]
\small
\setcounter{tempequationcounter}{\value{equation}}
\begin{IEEEeqnarray}{rCl}
\setcounter{equation}{21}
\hspace{-0.2in}\underbrace{\begin{bmatrix}
\bK(0)&   \cdots & \bK(\tau_1 -\tau_K) & \kappa \bK'(0) &   \cdots & \kappa \bK'(\tau_1 -\tau_K) \\
\vdots &\ddots &   \vdots & \vdots & \cdots & \vdots \\
\bK(\tau_K -\tau_1) &\cdots &   \bK(0) & \kappa \bK'(\tau_K -\tau_1) & \cdots & \kappa\bK'(0) \\
-\kappa\bK'(0) &   \cdots & -\kappa\bK'(\tau_1 -\tau_K) & -\kappa^2\bK''(0) &   \cdots & -\kappa^2\bK''(\tau_1 -\tau_K) \\
\vdots &\ddots &   \vdots & \vdots & \cdots & \vdots \\
-\kappa \bK'(\tau_K -\tau_1) &\cdots &   -\kappa \bK'(0) & -\kappa^2\bK''(\tau_K -\tau_1) & \cdots & -\kappa^2\bK''(0) 
\end{bmatrix}}_{\bGamma} \begin{bmatrix}
\balpha_1 \\
\vdots \\
\balpha_K \\
\kappa^{-1}\bbeta_1 \\
\vdots \\
\kappa^{-1}\bbeta_K 
\end{bmatrix} =  \begin{bmatrix}
\mbox{sign}(a_1^*) \bh^*  \\
\vdots \\
\mbox{sign}(a_K^*) \bh^*  \\
\bf{0} \\
\vdots \\
\bf{0}
\end{bmatrix}
\label{inversion}
\end{IEEEeqnarray}
\hrulefill
\begin{IEEEeqnarray}{rCl}
\setcounter{equation}{22}
\hspace{-0.15in}\bPhi &=\begin{bmatrix}
K(0) &   \cdots & K(\tau_1 -\tau_K) & \kappa K'(0) &   \cdots & \kappa K'(\tau_1 -\tau_K) \\
\vdots &\ddots &   \vdots & \vdots & \cdots & \vdots \\
K(\tau_K -\tau_1) &\cdots &   K(0) & \kappa K'(\tau_K -\tau_1) & \cdots & \kappa K'(0) \\
-\kappa K'(0) &   \cdots & -\kappa K'(\tau_1 -\tau_K) & -\kappa^2 K''(0) &   \cdots &  -\kappa^2 K''(\tau_1 -\tau_K) \\
\vdots &\ddots &   \vdots & \vdots & \cdots & \vdots \\
-\kappa K'(\tau_K -\tau_1) &\cdots &   -\kappa K'(0) &  -\kappa^2 K''(\tau_K -\tau_1) & \cdots &  -\kappa^2 K''(0) 
\end{bmatrix}  = \frac{1}{M} \sum_{n=-2M}^{2M} s_n \bnu_n \bnu_n^H
\label{Phi_definition}
\end{IEEEeqnarray}
\setcounter{equation}{\value{tempequationcounter}}
\hrulefill 
\normalsize
\end{figure*}

\subsection{Construction of the dual polynomial}\label{sec:construction_dual}
Without loss of generality, we assume $\| \bh\|_2 = 1$ from now on. Consider the squared Fejer's kernel and its derivative as
\begin{align*}
K(\tau) &=\frac{1}{M} \sum_{n=-2M}^{2M} s_n e^{-j2\pi \tau n}, \\ 
K'(\tau) &=\frac{1}{M} \sum_{n=-2M}^{2M} s_n (-j2\pi n) e^{-j2\pi \tau n}.
\end{align*}
where $s_{n}=\frac{1}{M}\sum_{i=\max\left(n-M,-M\right)}^{\min\left(n+M,M\right)}\left(1-\left\vert\frac{i}{M}\right\vert\right)\left(1-\left\vert\frac{n-i}{M}\right\vert\right)$ following the definition \cite{candes2014towards}. To proceed, we define randomized matrix-valued versions of $K(\tau)$ and $K'(\tau)$ as
\begin{align*}
\bK(\tau)&= \frac{1}{M} \sum_{n=-2M}^{2M} s_n (\bb_n\bb_n^H)e^{- j2\pi  \tau n} \in \mathbb{C}^{L\times L}, \\
 \bK'(\tau) &= \frac{1}{M} \sum_{n=-2M}^{2M} s_n (-j2\pi n)(\bb_n\bb_n^H)e^{-j2\pi \tau n}\in \mathbb{C}^{L\times L}, 
\end{align*}
where the derivatives are entry-wise. Clearly,
$$ \mathbb{E}\bK(\tau) = \frac{1}{M} \sum_{n=-2M}^{2M} s_n \expectation(\bb_n\bb_n^H)e^{-j2\pi \tau n} = K(\tau) \bI_L, $$
and $\expectation\bK'(\tau) = K'(\tau)\bI_L$, where $\bI_L$ is the $L\times L$ identity matrix following the isotropy property. We then construct the vector-valued dual polynomial $\bQ(\tau)\in\mathbb{C}^{L}$ as
\begin{equation}\label{dual_certificate}
\bQ(\tau) =  \sum_{k=1}^K \bK(\tau-\tau_k) \balpha_k  +  \sum_{k=1}^K \bK'(\tau-\tau_k) \bbeta_k  ,
\end{equation}
where $\balpha_k=[\alpha_{k,1},\ldots, \alpha_{k,L}]^T\in\mathbb{C}^{L\times 1}$ and $\bbeta_k=[\beta_{k,1},\ldots, \beta_{k,L}]^T \in\mathbb{C}^{L\times 1}$, $k=1,\ldots, K$. \yc{It is straightforward to see that $\bQ(\tau)$ has the valid form defined in \eqref{form_of_Q} of Proposition~\ref{prop_optimality}.} We select the coefficient vectors as the solution to the linear equations given below
\begin{equation}
\left\{ \begin{array}{lc}
\bQ(\tau_k) = \mbox{sign}(a_k^*)\bh^*, & \tau_k \in \mathcal{T} , \\
\bQ'(\tau_k) =\boldsymbol{0}, & \tau_k \in \mathcal{T}. \end{array} \right.
\end{equation}
which can be rewritten as equation \eqref{inversion}%
\addtocounter{equation}{1} after denoting $\kappa = 1/\sqrt{|K^{''}(0)|}$. For simplicity, we denote the LHS matrix of \eqref{inversion} as $\bGamma \in \mathbb{C}^{2LK\times 2LK}$. Before inverting \eqref{inversion}, we need to establish that $\bGamma$ is invertible with high probability.

\subsection{Invertibility of $\bGamma$}
The expectation of $\bGamma$ can be given as $\expectation\bGamma = \bar{\bGamma}  = \bPhi \kron \bI_L$, where $\kron$ is the Kronecker product, $\bPhi\in\mathbb{C}^{2K\times 2K}$ is given in \eqref{Phi_definition}%
\addtocounter{equation}{1}, where 
$$ \bnu_n = \begin{bmatrix}
e^{-j2\pi \tau_1 n} \\
\vdots \\
e^{-j2\pi \tau_K n} \\
(j2\pi \yc{n} \kappa) e^{-j2\pi \tau_1 n} \\
\vdots \\
(j2\pi \yc{n} \kappa) e^{-j2\pi \tau_K n} 
\end{bmatrix}. $$
The following lemma is useful from \cite{tang2012compressed} regarding $\bPhi$.
\begin{lemma}\cite[Proposition IV.1]{tang2012compressed}
\label{lemma_phi}
Let $\Delta\geq 1/M$. Then $\bPhi$ is invertible and
\begin{align*}
\left\Vert \boldsymbol{I}- {\bPhi} \right\Vert & \leq 0.3623, \;
\left\Vert{\bPhi} \right\Vert  \leq 1.3623, \;
\left\Vert{\bPhi}^{-1} \right\Vert  \leq 1.568.
\end{align*}
\end{lemma}
Note that we can write $\bGamma$ as a sum of independent random matrices as
\begin{align*}
\bGamma  & =  \frac{1}{M} \sum_{n=-2M}^{2M} s_n (\bnu_n\kron \bb_n) (\bnu_n\kron \bb_n)^H \\
& =  \frac{1}{M} \sum_{n=-2M}^{2M} s_n (\bnu_n \bnu_n^H) \kron (\bb_n\bb_n^H),
\end{align*}
then $\bGamma - \bar{\bGamma}$ can be written as
\begin{align} \label{deviation}
\bGamma - \bar{\bGamma} & = \sum_{n=-2M}^{2M} \bS_n,
\end{align}
where $\bS_n =  \frac{1}{M}  s_n (\bnu_n \bnu_n^H) \kron (\bb_n\bb_n^H - \bI_L) \in\mathbb{C}^{2KL\times 2KL}$.
The following lemma, whose proof is given in Appendix~\ref{proof_concentration},  establishes the concentration of $\bGamma$ around $\bar{\bGamma}$. 
\begin{lemma}\label{prop_concentration}
Let $0<\delta<1$ and $\Delta\geq 1/M$. For any $\chi\in(0, 0.6376)$, as long as
\begin{equation}\label{sample_concentraion}
M \geq \frac{80\mu KL}{\chi^2}\log\left(\frac{4KL}{\delta} \right),
\end{equation}
we have 
$ \left\Vert \bGamma - \bar{\bGamma}  \right\Vert \leq \chi $
holds with probability at least $1-\delta$. 
\end{lemma}
Denote the event $\mathcal{E}_{1,\chi}=\{\|\bGamma - \bar{\bGamma}  \| \leq \chi \}$, which holds with probability at least $1-\delta$ as long as \eqref{sample_concentraion} holds. This implies that the matrix $\bGamma$ is invertible when $\mathcal{E}_{1,\chi}$ holds for some $0<\chi<0.6376$, since
\begin{align*}
\left\Vert\boldsymbol{I}-\bGamma \right\Vert &\leq \left\Vert \boldsymbol{I}-\bar{\bGamma} \right\Vert+\left\Vert\bar{\bGamma}  - \bGamma \right\Vert \leq0.3623+\chi < 1,
\end{align*}
where $\left\Vert \boldsymbol{I}-\bar{\bGamma} \right\Vert=\left\Vert \boldsymbol{I}- {\bPhi} \right\Vert\leq 0.3623$ from Lemma~\ref{lemma_phi}. Under $\mathcal{E}_{1,\chi}$, let $ \bGamma^{-1} =  \begin{bmatrix}
{\bL} & {\bR} 
\end{bmatrix} $, where $\bL\in\mathbb{C}^{2LK\times LK}$ and $\bR\in\mathbb{C}^{2LK\times LK}$, then we have
\begin{align}
\hspace{-0.1in} \begin{bmatrix}
\balpha_1 \\
\vdots \\
\balpha_K \\
\kappa^{-1}\bbeta_1 \\
\vdots \\
\kappa^{-1}\bbeta_K 
\end{bmatrix} & = \bGamma^{-1} \begin{bmatrix}
\mbox{sign}(a_1^*) \bh^*  \\
\vdots \\
\mbox{sign}(a_K^*) \bh^*  \\
\bf{0} \\
\vdots \\
\bf{0}
\end{bmatrix} = \bL  [\mbox{sign}(\ba^*)\kron \bh^*]. \vspace{-0.1in} \label{coefficient_solution}
\end{align}
With the above parameterization, $\bQ(\tau)$ satisfies the first condition in \eqref{condition_one}. Let $\bPhi^{-1} = \begin{bmatrix}
\bar{\bL} & \bar{\bR} 
\end{bmatrix}$, where $\bar{\bL}\in\mathbb{C}^{2K\times K}$ and $\bar{\bR}\in\mathbb{C}^{2K\times K}$. Then we have
 $$ \bar{\bGamma}^{-1} = \bPhi^{-1}\kron \bI_L  = \begin{bmatrix}
\bar{\bL} & \bar{\bR} 
\end{bmatrix} \kron \bI_L =\begin{bmatrix}
\bar{\bL}\kron \bI_L & \bar{\bR} \kron \bI_L
\end{bmatrix},$$
and $\left\| \bar{\bGamma}^{-1} \right\|  = \| \bPhi^{-1}\|\leq 1.568$. Then using elementary linear algebra that is essentially the same as \cite[Corollary IV.5]{tang2012compressed}, we have the following lemma.
\begin{lemma}\label{lemma_gamma_inv}
On the event $\mathcal{E}_{1,\chi}$ with $\chi\in(0,1/4]$, we have
$$ \left\| \bGamma^{-1} -  \bar{\bGamma}^{-1} \right\| \leq 2 \left\| \bar{\bGamma}^{-1} \right\|^2 \chi, \quad  \left\| \bGamma^{-1} \right\| \leq 2 \left\|  \bar{\bGamma}^{-1} \right\| . \vspace{-0.05in}$$
\end{lemma}

\subsection{Certifying \eqref{condition_two}}
The rest of the proof is to guarantee $\bQ(\tau)$ satisfies \eqref{condition_two} with high probability. Let  $\bK^{(m)}(\tau)$ be the $m$th order entry-wise derivative of $\bK(\tau)$, $m=0,1,2,3$. The $(\ell,s)$th entry of $\bK^{(m)}(\tau)$ can be written as
$$ K_{\ell, s}^{(m)}(\tau) = \frac{1}{M} \sum_{n=-2M}^{2M} s_n(-j2\pi n)^{m} (b_{n,\ell}b_{n,s}^*)e^{- j2\pi  \tau n},$$  
where $b_{n,\ell}$ is the $\ell$th entry of $\bb_n$. Further define $\bXi^{(m)}(\tau)\in\mathbb{C}^{2LK\times L}$ as
\begin{align*}
\bXi^{(m)}(\tau) & = \kappa^{m}\begin{bmatrix}
\bK^{(m)}(\tau-\tau_1)^* \\
 \vdots \\
\bK^{(m)}(\tau-\tau_K)^* \\
\kappa \bK^{(m+1)}(\tau-\tau_1)^*\\ 
\vdots \\
\kappa  \bK^{(m+1)}(\tau-\tau_K)^*\\ 
\end{bmatrix} \\
& =  \frac{1}{M} \sum_{n=-2M}^{2M} s_n(j2\pi \kappa n)^{m}e^{j2\pi\tau n}(\bnu_n\kron \bb_n) \bb_n^H , 
\end{align*}
whose expectation $\bar{\bXi}^{(m)}(\tau) = \mathbb{E}\bXi^{(m)}(\tau) $ can be written as
\begin{align*}
\bar{\bXi}^{(m)}(\tau) & = \frac{1}{M} \sum_{n=-2M}^{2M} s_n(j2\pi \kappa n)^{m}e^{j2\pi\tau n} \left( \bnu_n\kron \bI_L \right) \\
&=\bzeta^{(m)}(\tau) \kron \bI_L ,  
\end{align*}
where 
\begin{align}
\bzeta^{(m)}(\tau) &= \kappa^{m} \begin{bmatrix}
K^{(m)}(\tau-\tau_1)^* \\
 \vdots \\
 K^{(m)}(\tau-\tau_K)^* \\
\kappa K^{(m+1)}(\tau-\tau_1)^*\\ 
\vdots \\
\kappa K^{(m+1)}(\tau-\tau_K)^*\\ 
\end{bmatrix} \nonumber \\
& =\frac{1}{M} \sum_{n=-2M}^{2M} s_n(j2\pi \kappa n)^{m} e^{j2\pi\tau n}\bnu_n. 
\end{align}

Let $\bQ^{(m)}(\tau)$ be the $m$th order entry-wise derivative of $\bQ(\tau)$, where the $m$th order derivative of the $\ell$th entry of $\bQ(\tau)$ can be written as
\small
$$Q_{\ell}^{(m)}(\tau) = \sum_{k=1}^K \sum_{s=1}^{L} \left[ K_{\ell,s}^{(m)}(\tau-\tau_k) \alpha_{k,s}  + K_{\ell,s}^{(m+1)}(\tau-\tau_k) \beta_{k,s}\right].$$
\normalsize
Then, $\bQ^{(m)}(\tau)$ can be written as 
\begin{align}\label{Qm_eq}
&\kappa^m\bQ^{(m)}(\tau)  = \left[\bXi^{(m)}(\tau)\right]^H \bL  [\mbox{sign}(\ba^*)\kron \bh^*] \nonumber \\
& =  \left[\bXi^{(m)}(\tau)-\bar{\bXi}^{(m)}(\tau) +\bar{\bXi}^{(m)}(\tau)  \right]^H \nonumber \\
& \quad\quad\quad \cdot\left( \bL - \bar{\bL}\kron \bI_L +\bar{\bL}\kron \bI_L \right)  [\mbox{sign}(\ba^*)\kron \bh^*] \nonumber\\
& = \left[\bar{\bXi}^{(m)}(\tau)  \right]^H \left(\bar{\bL}\kron \bI_L \right)   [\mbox{sign}(\ba^*)\kron \bh^*] \nonumber \\
& \quad\quad +\bI_1^{(m)}(\tau)+ \bI_2^{(m)}(\tau) , 
\end{align}
where
$$\bI_1^{(m)}(\tau) = \left[\bXi^{(m)}(\tau)-\bar{\bXi}^{(m)}(\tau)  \right]^H\bL  [\mbox{sign}(\ba^*)\kron \bh^*]$$
and
$$ \bI_2^{(m)}(\tau) = \left[\bar{\bXi}^{(m)}(\tau)  \right]^H\left( \bL - \bar{\bL}\kron \bI_L\right) [\mbox{sign}(\ba^*)\kron \bh^*].$$
Furthermore, the first term in \eqref{Qm_eq} can be written as
\begin{align*}
 &\quad \left[\bar{\bXi}^{(m)}(\tau)  \right]^H \left(\bar{\bL}\kron \bI_L \right)   [\mbox{sign}(\ba^*)\kron \bh^*]  \\
 & =( \bzeta^{(m)}(\tau)^H  \kron \bI_L) [\bar{\bL}\mbox{sign}(\ba^*)\kron \bh^*] \\
 & = \left[\bzeta^{(m)}(\tau)^H\bar{\bL}\mbox{sign}(\ba^*)\right] \bh^*:= \kappa^m \bar{Q}^{(m)}(\tau)\bh^*,
\end{align*}
where $\bar{Q}^{(m)}(\tau)= [\bzeta^{(m)}(\tau)]^H\bar{\bL}\mbox{sign}(\ba^*)$ becomes the scalar-valued dual polynomial constructed in \cite{candes2014towards}. Now, $\kappa^m\bQ^{(m)}(\tau) $ in \eqref{Qm_eq} can be rewritten as
\begin{equation}\label{dual_decomposition}
\yc{\kappa^m\bQ^{(m)}(\tau) = \kappa^m \bar{Q}^{(m)}(\tau)\bh^* + \bI_1^{(m)}(\tau) + \bI_2^{(m)}(\tau).}
\end{equation}

The rest of the proof proceeds in the following steps. We first establish that $\left\| \kappa^{m}\bQ^{\left(m\right)}\left(\tau\right)-\kappa^{m}\bar{Q}^{\left(m\right)}\left(\tau\right)\bh^*\right\|_2$ is uniformly bounded for points on a grid $\Upsilon_{\mathrm{grid}}$; next, we extend that  $\left\| \kappa^{m}\bQ^{\left(m\right)}\left(\tau\right)-\kappa^{m}\bar{Q}^{\left(m\right)}\left(\tau\right)\bh^*\right\|_2$ is uniformly bounded for all $\tau\in[0,1)$; finally, we show that $\left\|  \bQ\left(\tau\right) \right\|_2 < 1$, $\forall \tau\in [0,1]\backslash \mathcal{T}$.

To bound $\bI_1^{(m)}(\tau)$, the central argument is to bound $\left\| \bXi^{(m)}(\tau)-\bar{\bXi}^{(m)}(\tau)\right\|$ for a fixed $\tau \in[0,1)$, which is based on a significantly modified argument of \cite[Lemma IV.6]{tang2012compressed} whose proof is provided in Appendix~\ref{proof_bound_xi}.

\begin{lemma}\label{bound_xi}
Let $\Delta\geq 1/M$ and fix $\tau\in[0,1)$. Let
$$\bar{\sigma}_{m}^2 = 2^{4m+1}\frac{\mu L}{M}\max\left\{ 1, \yc{2KL\sqrt{\frac{2\mu }{M}} }\right\}, $$ 
and fix a positive number
$$ a \leq \left\{ \begin{array}{ll}
  \left(\frac{M}{8 \mu L}\right)^{1/4} & \quad \mbox{if}~  \yc{2KL\sqrt{\frac{2\mu }{M}} } \geq 1, \\
\sqrt{\frac{M}{8\mu KL}} & \quad \mbox{otherwise}.
\end{array}\right.  $$
then we have
\begin{align*}
 \left\| {\bXi}^{(m)}(\tau)  - \bar{\bXi}^{(m)}(\tau)\right\|  \leq   \yc{ 4^{m}L\sqrt{\frac{2\mu K}{M}}} + a \bar{\sigma}_m, 
\end{align*}
holds for $m=0,1,2,3$ with probability at least $1-  64e^{-ca^2}$ for some $c>0$.
\end{lemma}

We then establish that $\left\| \bI_1^{(m)}(\tau) \right\|_2$ is bounded on the grid $\Upsilon_{\mathrm{grid}}\in [0,1)$ in the following lemma, proved in Appendix~\ref{proof_bound_I1}.
\begin{lemma}\label{lemma_bound_I1}
Let $0<\delta<1$ and $\Delta\geq 1/M$. Let $\Upsilon_{\mathrm{grid}}$ be a finite set defined on $[0,1]$. As long as
\begin{multline}\label{sample_I1}
M \geq C \mu  \max\Big\{  \yc{ \frac{K^2L^2}{\epsilon^2} }  \log\left( \frac{64 |\Upsilon_{\mathrm{grid}}|}{\delta} \right), \\
  L \log^2 \left( \frac{64|\Upsilon_{\mathrm{grid}}|}{\delta} \right)  , KL \log\left(\frac{4KL}{\delta} \right) \Big\},
\end{multline}
for some constant $C$, we have 
$$\mathbb{P}\left\{ \sup_{\tau_d\in\Upsilon_{\mathrm{grid}}} \left\|\bI_{1}^{(m)}(\tau_d)\right\|_2 \leq  \epsilon, \; m=0,1,2,3 \right\} \geq 1- 16\delta.$$
\end{lemma}

We next bound $\bI_{2}(\tau)$, which is supplied in the following lemma proved in Appendix~\ref{proof_bound_l2}.
\begin{lemma}\label{lemma_bound_I2}
Let $0<\delta<1$ and $\Delta\geq 1/M$.  As long as
\begin{equation}\label{sample_I2}
M \geq C \mu K^2L \log\left(\frac{4KL}{\delta} \right),
\end{equation}
for some constant $C$, we have 
$$\mathbb{P}\left\{ \left\| \bI_{2}^{(m)}(\tau) \right\|_2\leq  \epsilon, \; m=0,1,2,3 \right\} \geq 1- 8\delta.$$
\end{lemma}

The next task is to extend the above inequalities to the unit interval $[0,1]$ by choosing the grid size properly. Define the event
\begin{align*}
\mathcal{E}_3 = \left\{ \left\| \kappa^m \bQ^{(m)}(\tau) -\kappa^{m}\bar{Q}^{\left(m\right)} \left(\tau\right) \bh^* \right\|_2   \leq \frac{\epsilon}{3},  m=0,1,2,3 \right\}.
\end{align*}
Combining Lemma~\ref{lemma_bound_I1} and Lemma~\ref{lemma_bound_I2}, with redefining the constants, we have the following lemma, which is proved in Appendix~\ref{proof_bound_Q}.
\begin{lemma} \label{lemma_bound_Q}
\yc{The event} $\mathcal{E}_3$ holds with probability at least $1-\delta$, as long as 
\begin{multline}
M \geq C  \mu \max\Big\{ \yc{\frac{K^2L^2}{\epsilon^2}}\log\left(\frac{M}{\epsilon\delta}\right)  \log\left(\frac{4KL}{\delta} \right), \\
 L\log^2 \left( \frac{M}{\epsilon\delta} \right)  , KL \log\left(\frac{4KL}{\delta} \right) \Big\} \label{requirement_lemma6}
\end{multline}
for some constant $C$.
\end{lemma}

Finally, we're ready to certify \eqref{condition_two}. Define a small neighborhood of each spike location as $T_i= \{ \tau: |\tau-\tau_i| \leq \rho_{c}/M\}$, where $\rho_c = 0.08245$. Let $T_{\mathrm{near}} = \cup_{i=1}^K T_i$ and $T_{\mathrm{far}} =  [0,1]\backslash T_{\mathrm{near}} $. We will establish the boundedness by splitting the analysis for $T_{\mathrm{near}}$ and $T_{\mathrm{far}}$. In fact, we have a stronger result in the following lemma, proved in Appendix~\ref{proof_far_near}. 
\begin{lemma}\label{bound_far_near}
Let $\Delta\geq 1/M$. We have 
\begin{subequations}
\begin{align}
\|  {\bQ}(\tau) \|_2 &   < 1- C_a, \quad  \forall \tau \in T_{\mathrm{far}}, \label{condition_far} \\
\left\|\bQ(\tau) \right\|_2   &  \leq  1 - C_bM^2(\tau-\tau_i)^2 , \quad \forall \tau\in T_i, \label{condition_near}
\end{align}
\end{subequations}
for some constants $C_a$ and $C_b$ satisfying $\rho_c^2C_b\leq C_a$, holds with probability at least $1-\delta$, as long as
$$ M \geq C  \mu  K^2 \yc{L^2}  \log^2\left(\frac{M}{ \delta}\right) $$
for some large enough constant $C$.
\end{lemma}
Putting all these together, we have now proved Theorem~\ref{main_theorem} since $\bQ(\tau)$ is verified to be a valid dual certificate.

\section{Conclusions}\label{conclusions}
This paper proposes a convex optimization framework for blind spikes deconvolution based on minimizing the atomic norm of a jointly spectrally-sparse ensemble after lifting the bilinear inverse problem into an under-determined linear inverse problem, by constraining the PSF to be in a known low-dimensional subspace. Under mild conditions, the proposed AtomicLift algorithm provably recovers the spike signal as well as the PSF up to a scaling ambiguity as long as the number of measurements is large enough.

Several interesting questions are left for future investigations. First, the phase transitions of AtomicLift suggests that it succeeds  when the number of measurements is $O(KL)$, which is better than what the theory guarantees, that is $O(K^2L^2)$ up to logarithmic factors. We believe it is possible to improve the performance guarantee of AtomicLift, for example by introducing further randomness assumptions on the spike signal \cite{tang2012compressed}. Second, the number of required measurements by AtomicLift is much larger than the degrees of freedom, which is on the order of $O(K+L)$. This may due to the fact that convex relaxations are not effective at exploiting joint structures in $\bZ^{\star}$ \cite{oymak2015simultaneously}. Indeed, the rank-one property of $\bZ^{\star}$ is not exploited in AtomicLift. A promising direction is to develop non-convex algorithms for blind spikes deconvolution, where recent work by Lee et. al. \cite{lee2015stability} could shed some light.
 
\section*{Acknowledgment}
The author thanks Louis L. Scharf for motivation to work on the problem of blind spikes deconvolution, and Zhi Tian, Yuxin Chen and Gongguo Tang for useful discussions. The author also gratefully acknowledges the anonymous reviewers for their constructive feedbacks that greatly improve the quality of this paper.

\appendices

\section{Proof of Proposition~\ref{prop_optimality}} \label{proof_optimality}

\begin{proof}
First, any $\bq$ satisfying \eqref{condition_one} and \eqref{condition_two} is dual feasible. We have
\begin{align*}
\|\bZ^{\star}\|_{\cA}  &  \geq\|\bZ^{\star}\|_{\cA}\|\mathcal{X}^*(\bq)\|_{\cA}^{\star} 
\geq \left\langle \mathcal{X}^*(\bq), \bZ^{\star} \right\rangle_{\mathbb{R}} \\
 &=\left\langle \mathcal{X}^*(\bq),  \sum_{k=1}^K a_k \bc(\tau_k)\bh^T \right\rangle_{\mathbb{R}} \\
& = \sum_{k=1}^K \mbox{Re}\left( a_k^* \left\langle \mathcal{X}^*(\bq), \bc(\tau_k)\bh^T \right\rangle \right) \\
& = \sum_{k=1}^K \mbox{Re} \left( a_k^* \langle \bQ^H(\tau_k), \bh^T \rangle \right) \\
& =  \sum_{k=1}^K \mbox{Re} \left( a_k^* \mbox{sign}(a_k)   \right)  =  \sum_{k=1}^K |a_k|  \geq \|\bZ^{\star}\|_{\cA} .
\end{align*}
Hence $\langle \mathcal{X}^*(\bq), \bZ^{\star} \rangle_{\mathbb{R}} =\|\bZ^{\star}\|_{\cA}$. By strong duality we have $\bZ^{\star}$ is primal optimal and $\bq$ is dual optimal.

For uniqueness, suppose $\hat{\bZ}$ is another optimal solution. If $\hat{\bZ}$ and $\bZ^{\star}$ have the same support set $\mathcal{T}$, they must coincide since the set of atoms in $\mathcal{T}$ is independent. Let $\hat{\bZ}=\sum_k \hat{a}_k \bc(\hat{\tau}_k)\hat{\bh}_k^T$ be its atomic decomposition where $\hat{a}_k >0$, with some support $\hat{\tau}_k \notin \mathcal{T}$. We then have 
\begin{align*}
&\quad \langle\mathcal{X}^*(\bq), \hat{\bZ}  \rangle_{\mathbb{R}}  \\
& = \sum_{\hat{\tau}_k\in\mathcal{T}}\mbox{Re}  \left( \hat{a}_k^* \langle  \bQ^H(\hat{\tau}_k) , \hat{\bh}_k^T \rangle \right) +  \sum_{\hat{\tau}_l\notin\mathcal{T}}\mbox{Re}  \left( \hat{a}_l^* \langle  \bQ^H(\hat{\tau}_l) , \hat{\bh}_l^T \rangle \right) \\
& \leq  \sum_{\hat{\tau}_k\in\mathcal{T}} \hat{a}_k \| \bQ(\hat{\tau}_k)\|_2 \|\hat{\bh}_k\|_2  + \sum_{\hat{\tau}_l \notin\mathcal{T}} \hat{a}_l  \| \bQ(\hat{\tau}_l)\|_2   \|\hat{\bh}_l\|_2 \\
& <\sum_{\hat{\tau}_k\in\mathcal{T}} \hat{a}_k   \|\hat{\bh}_k\|_2  + \sum_{\hat{\tau}_l \notin\mathcal{T}} \hat{a}_l      \|\hat{\bh}_l\|_2    = \|\hat{\bZ}\|_{\cA},
\end{align*}
which contradicts strong duality. Therefore the optimal solution of \eqref{atomic_lifting} is unique.
\end{proof}

\section{Proof of Proposition~\ref{prop_concentration}} \label{proof_concentration}
\begin{proof}
We apply the non-commutative Bernstein's inequality \cite{tropp2012user} to \eqref{deviation}. We have $\expectation\bS_n = 0$, and 
\begin{align*}
\left\| \bS_n \right\|  & =\frac{1}{M} |s_n| \cdot \left\| \bnu_n \bnu_n^H \right\| \cdot  \left\| \bb_n\bb_n^H - \bI_L \right\| \\
& \leq \frac{1}{M} \max_n |s_n| \cdot \| \bnu_n\|_2^2 \cdot \max\{ \|\bb_n\|_2^2 , \| \bI_L\| \} \\
& \leq \frac{1}{M} \cdot \left(K +K  \left(2\pi n\kappa\right)^{2} \right) \max\{ \mu L , 1 \} \\
& \leq \frac{14K  \mu L}{M} :=R.
\end{align*}
where in the first inequality we used that for two positive semidefinite matrices $\bA$ and $\bB$, $\| \bA - \bB\| \leq \max\{\|\bA\|, \|\bB\| \}$, in the second inequality we used that $ \max_n |s_n| \leq 1$ and the incoherence property of $F$, in the third inequality we used that $\left(1+\max_{\left\vert n\right\vert\le2M} \left(2\pi n\kappa\right)^{2} \right)\leq  14$ for $M\ge 4$ \cite{candes2014towards}. Moreover,
\begin{align*}
&\quad \left\| \sum_{n=-2M}^{2M} \mathbb{E}[\bS_n\bS_n^H] \right\| \\
 & =\frac{1}{M^2} \Big\| \sum_{n=-2M}^{2M} \mathbb{E}\Big\{ s_n^2 \left[(\bnu_n \bnu_n^H) \kron (\bb_n\bb_n^H - \bI_L)\right]  \\
 &\quad\quad\quad\quad\quad \cdot\left[(\bnu_n \bnu_n^H) \kron (\bb_n\bb_n^H - \bI_L)\right]^H \Big\} \Big\| \\
& =  \frac{1}{M^2}  \Big\| \sum_{n=-2M}^{2M} s_n^2 \|\bnu_n\|_2^2(\bnu_n \bnu_n^H) \\   &\quad\quad\quad\quad\quad \kron\mathbb{E}\left[  (\bb_n\bb_n^H - \bI_L) (\bb_n\bb_n^H - \bI_L)\right] \Big\| \\
& =  \frac{1}{M^2}  \left\| \sum_{n=-2M}^{2M} s_n^2 \|\bnu_n\|_2^2(\bnu_n \bnu_n^H)   \kron\mathbb{E}\left(\|\bb_n\|_2^2\bb_n\bb_n^H - \bI_L\right) \right\| \\
& \leq   \frac{14K \mu L}{M^2} \left\| \sum_{n=-2M}^{2M}s_n^2  (\bnu_n \bnu_n^H) \kron \bI_L \right\| \\
& \leq \frac{14K \mu L}{M} \max_n |s_n|  \left\| \frac{1}{M}\sum_{n=-2M}^{2M} s_n  (\bnu_n \bnu_n^H) \kron \bI_L \right\| \\
& \leq \frac{14K \mu L}{M} \left\| \bar{\bGamma}\right\| \leq  \frac{20K \mu L}{M}:=\sigma^2.
\end{align*}
where we used $ \|\bnu_n\|_2^2(\bnu_n \bnu_n^H)\preceq 14K \bnu_n \bnu_n^H$, $\|\bb_n\|_2^2\bb_n\bb_n^H\preceq \mu L \bb_n\bb_n^H$, and $\left\| \bar{\bGamma}\right\| = \left\|\bPhi \right\| \leq 1.3623$ from Lemma~\ref{lemma_phi}. The notation $\bA\preceq \bB$ indicates $\bB-\bA$ is a positive semidefinite matrix. Applying Bernstein's inequality to \eqref{deviation} will then finish the proof.
\end{proof}

\section{Proof of Lemma~\ref{bound_xi}}\label{proof_bound_xi}
\begin{proof}
We first write  
\begin{align*}
&\quad {\bXi}^{(m)}(\tau)  - \bar{\bXi}^{(m)}(\tau)  = \sum_{n=-2M}^{2M} \bW_n^{(m)} \\
& = \frac{1}{M} \sum_{n=-2M}^{2M} s_n(j2\pi \kappa n)^{m}e^{j2\pi\tau n}\bnu_n\kron( \bb_n \bb_n^H - \bI_L ) 
\end{align*}
where $\bW_n^{(m)} = \frac{1}{M}s_n(j2\pi \kappa n)^{m}e^{j2\pi\tau n}\bnu_n\kron( \bb_n \bb_n^H - \bI_L ) \in\mathbb{C}^{2LK\times L}$'s are independent random matrices with zero mean. Define
\begin{align*}
 V_m &= \left\|{\bXi}^{(m)}(\tau)  - \bar{\bXi}^{(m)}(\tau)\right\| \\
& = \sup_{\|\bu\|_2=1,\|\bv\|_2=1} \mbox{Re}\left[ \bu^H\left({\bXi}^{(m)}(\tau)  - \bar{\bXi}^{(m)}(\tau) \right)\bv \right] \\
& =  \sup_{  \|\bu\|_2=1,\|\bv\|_2=1}  \sum_{n=-2M}^{2M} \mbox{Re}\left( \bu^H\bW_n^{(m)} \bv \right) 
\end{align*}
We apply Talagrand's concentration inequality in Lemma~\ref{talagrand} to bound $V_m$.
\begin{lemma} \cite[Talagrand's concentration inequality]{talagrand1995concentration}  \label{talagrand}
Let $\{Y_j\}$ be a finite sequence of independent random variables taking values in a Banach space and $V$ be defined as
$ V = \sup_{ h\in\mathcal{H}} \sum_j h (Y_j) $
for a countable family of real valued functions $\mathcal{H}$. Assume that $|h|\leq B$ and $\mathbb{E}h (Y_j) = 0$ for all $h\in \mathcal{H}$ and every $j$. Then for all $t > 0$,
$$\mathbb{P} \{ |V - \mathbb{E}V | > t \} \leq 16\exp\left( -\frac{t}{KB}\log\left( 1+ \frac{Bt}{\sigma^2+B\mathbb{E} \bar{V}}\right)\right),$$
where $\sigma^2= \sup_{h\in\mathcal{H}} \sum_j \mathbb{E}h^2(Y_j) $, $\bar{V}=\sup_{h\in\mathcal{H}} \left| \sum_j h(Y_j) \right| $, and $K$ is a numerical constant.
\end{lemma}

Let $ h(\bW_n^{(m)})= \mbox{Re}\left( \bu^H\bW_n^{(m)} \bv \right)$, then $\mathbb{E}h(\bW_n^{(m)}) =0$. We compute the following bounds:
\begin{align*}
& \quad | h(\bW_n^{(m)})| \\
&= \left|  \mbox{Re}\left[ \bu^H\left( \frac{s_n}{M} (j2\pi \kappa n)^{m}e^{j2\pi\tau n}\bnu_n\kron( \bb_n \bb_n^H - \bI_L ) \right)\bv \right] \right| \\
& \leq  4^{m+1}\sqrt{K}\frac{\mu L}{M}: =B_{m} ,
\end{align*}
where we have used $ \max_n |s_n| \leq 1$, $|j2\pi \kappa n|\leq 4$, and $\|\bnu_n\|_2\leq  \sqrt{14K}$ for $M\ge 4$ \cite{candes2014towards}, and $\|\bb_n\|_2^2\leq \mu L$. Since
\begin{align*}
&\quad \mathbb{E} \left\|{\bXi}^{(m)}(\tau)  - \bar{\bXi}^{(m)}(\tau)\right\|^2  \leq \mathbb{E} \left\|{\bXi}^{(m)}(\tau)  - \bar{\bXi}^{(m)}(\tau)\right\|_{\mathrm{F}}^2  \\
& = \mathbb{E} \mbox{Tr} \left(\sum_{n=-2M}^{2M} \bW_n^{(m)}\right) \left(\sum_{n'=-2M}^{2M} \bW_{n'}^{(m)}\right)^H  \\
& = \mbox{Tr}  \mathbb{E} \left(\sum_{n=-2M}^{2M} \bW_n^{(m)}(\bW_n^{(m)})^H \right) \\
& = \frac{1}{M^2 }\sum_{n=-2M}^{2M} |s_n|^2 (2\pi\kappa n)^{2m} \mbox{Tr}(\bnu_n\bnu_n^H)\cdot \mbox{Tr}(\mathbb{E}(\bb_n\bb_n^H -\bI_L)^2) \\
& \leq  \frac{4^{2m}}{M^2 }\mbox{Tr}\left(\sum_{n=-2M}^{2M} s_n \bnu_n\bnu_n^H \right) \mu \yc{L^2} \\
& \leq \frac{4^{2m}}{M} \mbox{Tr}\left(\bPhi\right) \mu \yc{L^2} \leq 2^{4m+1}\frac{\mu K\yc{L^2}}{M},
\end{align*}
where in the first inequality we used $\|\bA\|\leq \|\bA\|_{\mathrm{F}}$, followed by exchanging the order of taking expectation and the trace and using the independence of $\bW_n^{(m)}$. In the second inequality, we again used $ \max_n |s_n| \leq 1$, $|j2\pi \kappa n|\leq 4$, and $\|\bnu_n\|_2\leq  \sqrt{14K}$ for $M\ge 4$ \cite{candes2014towards}, and $\|\bb_n\|_2^2\leq \mu L$. The last inequality follows from $\mbox{Tr}(\bPhi)=2K$. This yields 
$$\mathbb{E}\bar{V}_m = \mathbb{E}{V}_m \leq \left( \mathbb{E}{V}_m^2\right)^{1/2}\leq 4^{m}\yc{L}\sqrt{\frac{2\mu K}{M}}.$$

Next, we have
\begin{align*}
& \mathbb{E} h^2(\bW_n^{(m)})  \leq \mathbb{E}\left\| \bu^H\bW_n^{(m)} \bv\right\|_2^2  \\
& = \frac{1}{M^2} |s_n|^2 (2\pi \kappa n)^{2m} \bu^H \\
& \quad\quad \cdot \left[ \bnu_n\bnu_n^H\kron \mathbb{E}(( \bb_n \bb_n^H - \bI_L ) \bv\bv^H( \bb_n \bb_n^H - \bI_L )) \right] \bu \\
& \leq  \frac{1}{M^2} |s_n|^2 (2\pi \kappa n)^{2m} \bu^H  \left[ \bnu_n\bnu_n^H\kron \mu L\bI_L \right] \bu \\
& \leq  \frac{\mu L}{M^2} 4^{2m}\bu^H  \left[ s_n\bnu_n\bnu_n^H\kron  \bI_L \right] \bu
\end{align*}
where in the first inequality we used
\begin{align*}
&\quad\mathbb{E}(( \bb_n \bb_n^H - \bI_L ) \bv\bv^H( \bb_n \bb_n^H - \bI_L ))  \\
&=\mathbb{E} (|\bb_n^H\bv|^2\bb_n\bb_n^H) -\bv\bv^H \\
& \preceq \|\bb_n \|_2^2  \bI_L  -\bv\bv^H \preceq \mu L  \bI_L.
\end{align*}
Therefore,
\begin{align*}
\sigma_m^2& =\sum_{n=-2M}^{2M} \mathbb{E} h^2(\bW_n^{(m)}) \\
& =  \frac{\mu L}{M^2} 4^{2m} \bu^H  \left[\sum_{n=-2M}^{2M} s_n\bnu_n\bnu_n^H\kron  \bI_L \right] \bu \\
& \leq  \frac{\mu L}{M} 4^{2m}  \| \bar{\bGamma}\|  \leq \frac{\mu L}{M}2^{4m+1},
 \end{align*}
where in the last inequality we used $\| \bar{\bGamma}\|\leq 2$. Applying Lemma~\ref{talagrand} we have
\begin{align*}
&  \mathbb{P} \left\{ \left|  \left\|{\bXi}^{(m)}(\tau)  - \bar{\bXi}^{(m)}(\tau)\right\|  - \mathbb{E} \left\|{\bXi}^{(m)}(\tau)  - \bar{\bXi}^{(m)}(\tau)\right\| \right| > t \right\} \\
& \leq 16\exp\left( -\frac{t}{CB_m}\log\left( 1+ \frac{B_m t}{\sigma_m^2+B_m\mathbb{E} \bar{V}_m}\right)\right)
\end{align*} 
for some constant $C$. Suppose 
$$\bar{\sigma}_{m}^2 = \max(\sigma_{m}^2, B_m\mathbb{E} \bar{V}_m) = 2^{4m+1}\frac{\mu L}{M}\max\left\{ 1, 2K\yc{L}\sqrt{\frac{2\mu }{M}} \right\}, $$ 
and fix $a$ obeying $a\leq \bar{\sigma}_m/B_m$,  we have
\begin{align*}
 \mathbb{P} \left\{  \left\| {\bXi}^{(m)}(\tau)  - \bar{\bXi}^{(m)}(\tau)\right\|  \geq    4^{m}\yc{L}\sqrt{\frac{2\mu K}{M}} + a \bar{\sigma}_m \right\} \leq 16e^{-ca^2}
\end{align*}
for some constant $c>0$, which finishes the proof.
\end{proof}

\section{Proof of Lemma~\ref{lemma_bound_I1}} \label{proof_bound_I1}
\begin{proof}
Define the event
\begin{multline*}
\mathcal{E}_2=\bigcap_{\tau_d\in\Upsilon_{\mathrm{grid}}}\Big\{ \left\|\bXi^{(m)}(\tau_d)-\bar{\bXi}^{(m)}(\tau_d)  \right\|  \\
\leq 2^{2m}\yc{L}\sqrt{\frac{2\mu K}{M}} + a\bar{\sigma}_m : =\lambda_m \Big\}. 
\end{multline*}
Conditioned on $\mathcal{E}_{1,\chi} \cap \mathcal{E}_2 $, we have
\begin{align*} 
\left\|\bI_{1}^{(m)}(\tau) \right\|_2 & \leq \left\|  \bXi^{(m)}(\tau)-\bar{\bXi}^{(m)}(\tau)  \right\| \cdot \|\bL\| \cdot\| \mbox{sign}(\ba^*)\kron \bh^*\| \\
& \leq 4\sqrt{K}\lambda_m ,
\end{align*}
where we used $\| \bL_k\|\leq \|\bL\|\leq  2\left\|  \bar{\bGamma}^{-1} \right\|\leq 4$ from Lemma~\ref{lemma_gamma_inv}. Set $\lambda_m \leq \epsilon/(4\sqrt{K})$. Applying the union bound and getting rid of the conditional probability we have
\begin{equation}
 \mathbb{P}\left(\sup_{\tau_d\in\Upsilon_{\mathrm{grid}}} \left\|\bI_{1}^{(m)}(\tau) \right\|_2  >\epsilon \right)  \leq 64  \left|\Upsilon_{\mathrm{grid}}\right|  e^{-c a^2} + \mathbb{P}(\mathcal{E}_{1,\chi}^c), \label{bound_tail}
\end{equation}
where the second term $ \mathbb{P}(\mathcal{E}_{1,\chi}^c)\leq \delta$ when $M \geq \frac{80\mu KL}{\chi^2}\log\left(\frac{4KL}{\delta} \right)$. Set
$ a^2 = c^{-1}\log\left( \frac{64|\Upsilon_{\mathrm{grid}}|}{\delta} \right) $ so the first term in \eqref{bound_tail} is also bounded by $\delta$.

If $2K\yc{L}\sqrt{\frac{2 \mu   }{M}}\geq 1$, from Lemma~\ref{bound_xi} we have $a\leq  \left(\frac{M}{8 \mu L}\right)^{1/4} $ which gives
$ M \geq \frac{64}{c^2} \mu L\log^2 \left( \frac{64|\Upsilon_{\mathrm{grid}}|}{\delta} \right)$.
Moreover, since $a\bar{\sigma}_m <2^{2m}\yc{L}\sqrt{\frac{2\mu K}{M}}$, we have $ \frac{1}{\lambda_m}>\frac{1}{2^{2m+1}\yc{L}}\sqrt{\frac{M}{2\mu K}}$. Then \eqref{bound_tail} holds as long as $\frac{1}{2^{2m+1}\yc{L}}\sqrt{\frac{M}{2\mu K}} \geq \frac{4\sqrt{K}}{\epsilon}$, which gives
$ M \geq 4^{2m+7} \frac{\mu K^2\yc{L^2}}{\epsilon^2}$. If $2K\yc{L}\sqrt{\frac{2 \mu   }{M}}< 1$, from Lemma~\ref{bound_xi} we have $a\leq \sqrt{\frac{M}{8\mu KL}}$ which gives $M \geq \frac{8}{c} \mu KL\log \left( \frac{64|\Upsilon_{\mathrm{grid}}|}{\delta} \right)$. The rest follows similarly as in the previous case. Set $\chi=1/4$, the proof is complete by combining all lower bounds on $M$ and absorbing the constants. 
\end{proof}

\vspace{-0.1in}
\section{Proof of Lemma~\ref{lemma_bound_I2}} \label{proof_bound_l2}
\begin{proof}
To bound $\bI_{2}^{(m)}(\tau)$, we recall from \cite{tang2012compressed} that
$ \|\bzeta^{(m)}(\tau)\|_2  \leq C_1 $
for some universal constant $C$, hence $ \left\| \bar{\bXi}^{(m)}(\tau)\right\| =\left\| \bzeta^{(m)}(\tau) \otimes \bI_L \right\| \leq C_1$. Moreover, conditioned on $\mathcal{E}_{1,\chi}$ with $\chi\in[0,1/4)$, 
\begin{align*}
\left\|\bI_2^{(m)}(\tau) \right\|_2 & \leq  \left\| \bar{\bXi}^{(m)}(\tau)\right\| \cdot \left\| \bL - \bar{\bL}\kron \bI_L\right\| \cdot \|\mbox{sign}(\ba^*)\kron \bh^*\|_2 \\
& \leq  C'\sqrt{K}\chi,
\end{align*}
To bound  this by $\epsilon$, set $\chi = \frac{\epsilon}{C'\sqrt{K}}$. Plug this into \eqref{sample_concentraion}, we require
$ M \geq C \mu K^2L \log\left(\frac{4KL}{\delta} \right)$
for some large enough constant $C$. 
\end{proof}

\section{Proof of Lemma~\ref{lemma_bound_Q}} \label{proof_bound_Q}

\begin{proof}
First of all, we have
\begin{align*}
\kappa^{m}|Q_{\ell}^{(m)}(\tau)|  &  \leq  C\sqrt{KL} \| \bL\| \|\mbox{sign}(\ba^*)\kron \bh^* \| \\
& \leq C \sqrt{KL} \sqrt{K} =CK\sqrt{L},
\end{align*}
then for any fixed $\tau_1$, $\tau_2$, by the Bernstein's polynomial inequality, we have
\begin{align*}
&\quad \left| \kappa^{m}Q_{\ell}^{(m)}(\tau_1) -   \kappa^{m}Q_{\ell}^{(m)}(\tau_2)   \right|  \\
& \leq |e^{-j2\pi\tau_1}-e^{-j2\pi\tau_2}| \sup_{z=e^{-j2\pi \tau}} \left|\frac{d\kappa^{m}Q_{\ell}^{(m)}(z)}{dz} \right| \\
& \leq 4\pi |\tau_1 -\tau_2| 2M \sup_{\tau}\left| \kappa^{m}Q_{\ell}^{(m)}(\tau) \right| \\
& \leq CMKL, 
\end{align*}
for some constant $C$. Therefore, we have 
\begin{align*}
&\quad \left\| \kappa^{m}\bQ^{(m)}(\tau_1) -   \kappa^{m}\bQ^{(m)}(\tau_2)   \right\|_2  \\
& \leq \left(\sum_{\ell=1}^L \left| \kappa^{m}Q_{\ell}^{(m)}(\tau_1) -   \kappa^{m}Q_{\ell}^{(m)}(\tau_2)   \right|^2\right)^{1/2}\\
& \leq CMKL^{1.5} \leq CM^4, 
\end{align*}
we can select the grid size such that for any $\tau\in[0,1]$, there exists a point $\tau_d \in \Upsilon_{\mathrm{grid}}$ satisfying $|\tau-\tau_d|\leq \frac{\epsilon}{3CM^4}$. The grid size can be $3CM^4/\epsilon$. With this selection, we have
\begin{align*}
\left\| \kappa^{m}\bQ^{(m)}(\tau) -  \kappa^{m}\bar{Q}^{(m)}(\tau) \bh^*   \right\|_2  \leq \epsilon, \quad \forall \tau\in [0,1],
\end{align*}
as long as $M$ satisfies the requirement in \eqref{requirement_lemma6}.
\end{proof}

\section{Proof of Lemma~\ref{bound_far_near}} \label{proof_far_near}

We first record some useful properties of $\bar{Q}(\tau)$ below, where most of them are borrowed from \cite[Proposition IV.2]{tang2012compressed}.
\begin{lemma}
\label{lemma_Qm}
Assume $\Delta\geq 1/M$. Then, for $\tau\in T_{\mathrm{far}}$, $|\bar{Q}(\tau)|<0.99992$, 
and for $\tau\in T_{\mathrm{near}}$,
\begin{align*} 
&  \kappa^2 \left(\bar{Q}_R(\tau) \bar{Q}_R^{''}(\tau) + |\bar{Q}^{'}(\tau)|^2 +|\bar{Q}_I(\tau)| |\bar{Q}_I^{''}(\tau)| \right)  \leq -0.07865,  
\end{align*}
$\kappa^2 |\bar{Q}{''}(\tau)|  < 1.7068$, and $ \kappa |\bar{Q}{'}(\tau)| < 0.4346$ .
\end{lemma}
Note that the bound $\kappa^2 |\bar{Q}{''}(\tau)|  < 1.7068$ in Lemma~\ref{lemma_Qm} is new, which follows directly from rewriting \cite[equation (2.27)]{candes2014towards} to bound $\kappa^2 |\bar{Q}_R^{''}(\tau)| $, hence we omit the details.

\begin{proof} First assume $\tau\in T_{\mathrm{far}}$. Using Lemma~\ref{lemma_bound_Q} we have
\begin{align*}
\| \bQ(\tau)\|_2 & = \|  \bh^* \bar{Q}(\tau)\|_2 +\|\bQ(\tau) - \bh^* \bar{Q}(\tau)\|_2\\
& \leq \|\bh^* \bar{Q}(\tau)\|_2 +\epsilon  \leq 0.99992 + \epsilon<1,
\end{align*}
as long as $\epsilon<10^{-5}$.

Next assume $\tau \in T_{\mathrm{near}}$. Since our choice of the coefficients implies that
\begin{align*}
\frac{d\| \bQ(\tau)\|_2^2}{d\tau} \Big|_{\tau=\tau_k} &= \mbox{Re}\left( 2\bQ(\tau)^H \frac{d\bQ(\tau)}{d\tau}\right) \Big|_{\tau=\tau_k} \\
&=2\bQ_R^T(\tau)\bQ_R^{'}(\tau)+2\bQ_I^T(\tau)\bQ_I^{'}(\tau)=0,
\end{align*}
it is sufficient to establish $\frac{d^2\| \bQ(\tau)\|_2^2}{d\tau^2}  <0$ for $\tau\in T_{\mathrm{near}}$. First of all,
\begin{align*}
 \frac{1}{2}\frac{d^2\| \bQ(\tau)\|_2^2}{d\tau^2} & =\bQ_R^T(\tau)\bQ_R{''}(\tau)+\|\bQ_R{'}(\tau)\|_2^2\\
 & \quad\quad\quad+\bQ_I(\tau)^T\bQ_I{''}(\tau) +\|\bQ_I{'}(\tau)\|_2^2 \\
 & = \mbox{Re}\left( (\bQ{''}(\tau))^H\bQ(\tau) \right)+\|\bQ{'}(\tau)\|_2^2 .
\end{align*}
Since 
\begin{align*}
\|\kappa\bQ{'}(\tau)\|^2 &= \|\kappa\bQ{'}(\tau) - \kappa\bh^* \bar{Q}{'}(\tau)+\kappa\bh^* \bar{Q}{'}(\tau)\|^2  \\
& \leq \epsilon^2 + \kappa^2 |\bar{Q}{'}(\tau)|^2  +0.8692\epsilon
\end{align*}
and
\begin{align*}
&\quad \kappa^2 \mbox{Re}\left( (\bQ{''}(\tau))^H\bQ(\tau) \right) \\
& = \kappa^2 \mbox{Re}\left( (\bar{Q}{''}(\tau))^*\bar{Q}(\tau) \right) + \kappa^2 \mbox{Re}\left( (\bQ{''}(\tau)-\bh^*\bar{Q}{''}(\tau))^H\bQ(\tau) \right)   \\
& \quad +\kappa^2 \mbox{Re}\left( (\bh^*\bar{Q}{''}(\tau))^H(\bQ(\tau) -\bh^*\bar{Q}(\tau)) \right)   \\
& \leq  \kappa^2 \mbox{Re}\left( (\bar{Q}{''}(\tau))^*\bar{Q}(\tau) \right) + \epsilon(1+\epsilon) + \yc{1.7068}\epsilon, 
\end{align*}
therefore,
\begin{align*}
\frac{\kappa^2}{2}\frac{d^2\| \bQ(\tau)\|_2^2}{d\tau^2} & \leq \kappa^2 \mbox{Re}\left( (\bar{Q}{''}(\tau))^*\bar{Q}(\tau) \right) + \epsilon(1+\epsilon) \\
& \quad\quad + \yc{1.7068}\epsilon +\epsilon^2 + \kappa^2 |\bar{Q}{'}(\tau)|^2  +0.8692\epsilon \\
& \leq \kappa^2  (\bar{Q}_R(\tau) \bar{Q}_R{''}(\tau) + |\bar{Q}(\tau){'}|^2 \\
& \quad\quad +|\bar{Q}_I(\tau)| |\bar{Q}_I{''}(\tau)|  ) +2\epsilon^2 +\yc{3.576}\epsilon \\
& \leq -7.865\times 10^{-2}+2\epsilon^2 +\yc{3.576}\epsilon <0
\end{align*}
for a small enough numerical constant $\epsilon$. Plug the choice of $\epsilon$ into the sample complexity requirement of Lemma~\ref{lemma_bound_Q}, the proof is accomplished by keeping only the dominating terms.
\end{proof}

\bibliography{bibToeplitz,bibfileToeplitzPR,draft}
\bibliographystyle{IEEEtran}

\begin{IEEEbiography}[{\includegraphics[width=1in,height=1.25in,clip,keepaspectratio]{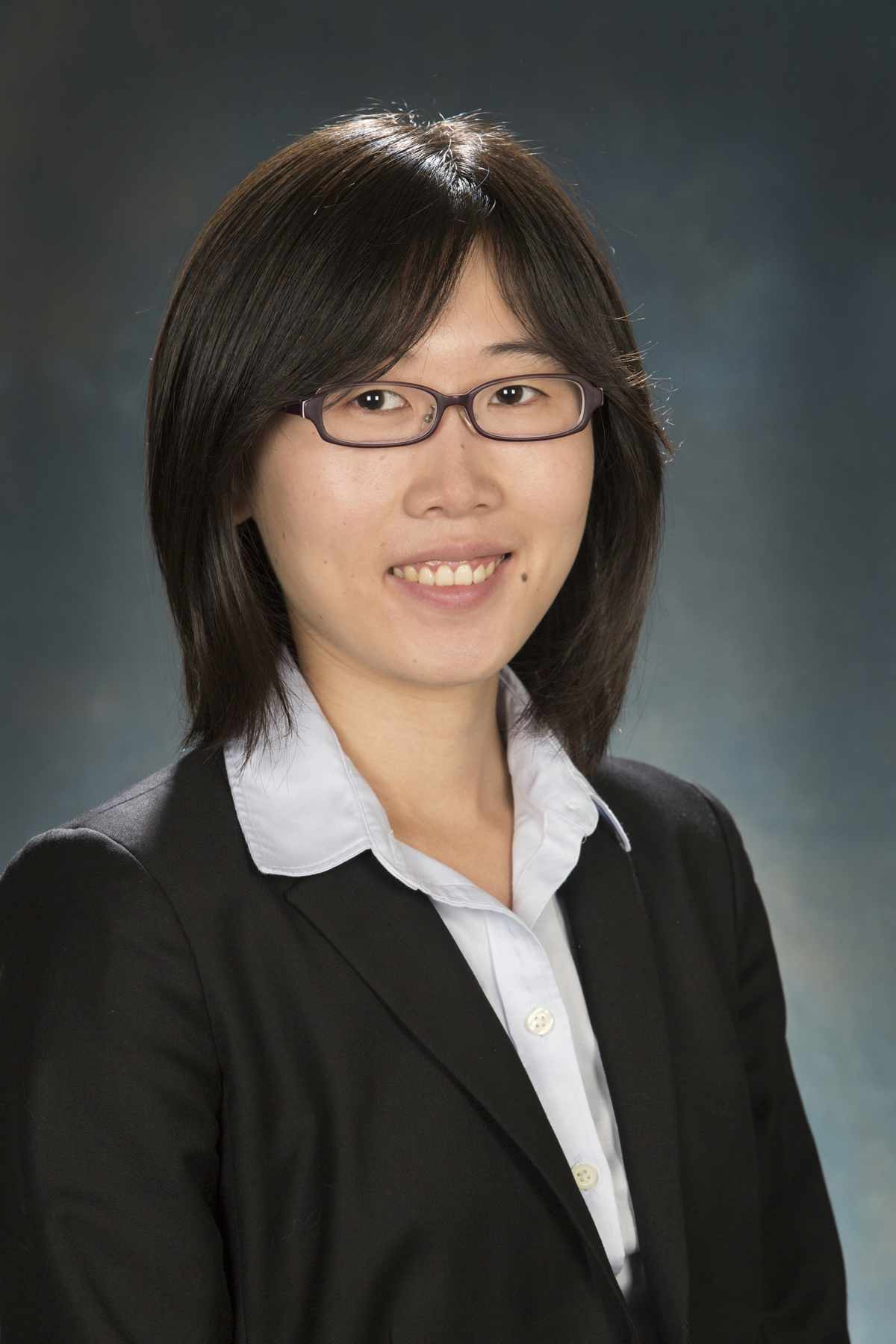}}]{Yuejie Chi}
(S'09-M'12) is an assistant professor with the department of Electrical and Computer Engineering and the department of Biomedical Informatics at The Ohio State University since September 2012. She received the Ph.D. degree in Electrical Engineering from Princeton University in 2012, and the B.E. (Hon.) degree in Electrical Engineering from Tsinghua University, Beijing, China.

She is the recipient of the IEEE Signal Processing Society Young Author Best Paper Award in 2013 and the Best Paper Award at the IEEE International Conference on Acoustics, Speech, and Signal Processing (ICASSP) in 2012. She received the Young Investigator Program Awards from AFOSR and ONR respectively in 2015, the Ralph E. Powe Junior Faculty Enhancement Award from Oak Ridge Associated Universities in 2014, a Google Faculty Research Award in 2013, the Roberto Padovani scholarship from Qualcomm Inc. in 2010. She is an Elected Member of the MLSP and SPTM Technical Committees of the IEEE Signal Processing Society since January 2016. She has held visiting positions at Colorado State University, Stanford University and Duke University, and interned at Qualcomm Inc. and Mitsubishi Electric Research Lab. Her research interests include statistical signal processing, information theory, machine learning and their applications in high-dimensional data analysis, network inference, active sensing and bioinformatics.

\end{IEEEbiography}

\end{document}